\documentclass[preprint]{aastex63}
\begin{document}
\pdfoutput=1

\pagestyle{plain}
\pagenumbering{arabic}
\parskip=5pt
\parindent=15pt

\newcommand{\ts}{\thinspace}
\newcommand{\about}{$\sim$\ts}
\newcommand{\etal}{et~al.}
\newcommand{\pad}{$p_{AD}$}

\title{Systematics in the ALMA Proposal Review Rankings}
\author[0000-0003-2251-0602]{John Carpenter}
\affiliation{Joint ALMA Observatory, Avenida Alonso de C\'ordova 3107, Vitacura, Santiago, Chile}

\begin{abstract} 

The results from the ALMA proposal peer review process in Cycles 0--6 are
analyzed to identify any systematics in the scientific rankings that
may signify bias. Proposal rankings are analyzed with respect to the
experience level of a Principal Investigator (PI) in submitting ALMA proposals,
regional affiliation (Chile, East Asia, Europe, North America, or
Other), and gender. The analysis was conducted for both the Stage 1 
rankings, which are based on the preliminary scores from the reviewers, and the
Stage 2 rankings, which are based on the final scores from the
reviewers after participating in a face-to-face panel discussion. Analysis of
the Stage 1 results shows that PIs who submit an ALMA proposal in multiple
cycles have systematically better proposal ranks than PIs who have submitted
proposals for the first time. In terms of regional affiliation, PIs from Europe
and North America have better Stage 1 rankings than PIs from Chile and East
Asia. Consistent with \citet{Lonsdale16}, proposals led by men have
better Stage 1 rankings than women when averaged over all cycles. This trend
was most noticeably present in Cycle 3, but no discernible differences in the
Stage 1 rankings are present in recent cycles. Nonetheless, in each
cycle to date, women have had a lower proposal acceptance rate than men even
after differences in demographics are considered. Comparison of the
Stage 1 and Stage 2 rankings reveal no significant changes in the distribution
of proposal ranks by experience level, regional affiliation, or gender as a
result of the panel discussions, although the proposal ranks for East Asian PIs
show a marginally significant improvement from Stage 1 to Stage 2 when averaged
over all cycles. Thus any systematics in the proposal rankings are introduced
primarily in the Stage 1 process and not from the face-to-face discussions.
These results are discussed in the context of potential language and cultural
biases, but any conclusions on the origin of the observed systematics remain
speculative.

\end{abstract}

\section{Introduction}

The Atacama Large Millimeter/Submillimeter Array (ALMA) is an international
astronomical facility operated in a partnership of the European Organisation
for Astronomical Research in the Southern Hemisphere (ESO), the U.S. National
Science Foundation, and the National Institutes of Natural Sciences of Japan in
cooperation with the Republic of Chile. The Joint ALMA Observatory (JAO)
solicits observing proposals from the scientific community to use ALMA through
an annual Call for Proposals. This is the primary means by which projects are
selected for observation. ALMA proposals are peer reviewed by volunteers from
the scientific community. Projects are added to the observing queue based
primarily on the scientific rank from the review process, but also
operational considerations, including over-subscription in antenna
configurations and by right ascension, the required weather conditions for the
observations, and the pre-determined share of observing time that is awarded 
to Chile, East Asia, Europe, and North America.

Given the importance that telescope access can have on developing a scientific
career, it is imperative that the community has confidence that scientific
merit is the primary determinant of proposal rank. Analyzing the results from
the review process and presenting the outcomes in a transparent manner is an
important part in building confidence within the community. Along these lines,
identifying potential systematics in the proposal review process at
astronomical observatories has received prominent attention in recent years. If
the probability of success of a proposal depends on some characteristic (e.g.,
the gender of the Principal Investigator, or PI) that should not correlate with
the underlying scientific merit, it may indicate a bias in the review process.

\citet{Reid14} raised attention to possible biases in the proposal review
process of the Hubble Space Telescope (HST) when he found that proposals led by
women had a lower acceptance rate than proposals led by men for HST Cycles 11
through 21. While the difference in acceptance rate by gender is not
significant in any given cycle, the persistent trend over time cannot be
attributed to random noise. \citet{Reid14} speculated on possible causes of the
gender-based systematic, but establishing the origin with any degree of
certainty was difficult. One possibility is that unconscious bias is present
among the reviewers that favor men over women in the scientific review.
However, the data contained hints that demographic differences may also play a
role. In response to these systematics, HST took the step of listing the
investigators alphabetically on the proposal cover sheet so that reviewers
could no longer identify the PI. Yet, women continued to have a lower
acceptance rate than men until a double-anonymous review was instituted that
hid the identity of all investigators to the reviewers \citep{Strolger19}.

Following the study by \citet{Reid14}, \citet{Patat16} analyzed the proposal
statistics for ESO and found that women also have had a lower success rate than
men when applying for observing time. He found that the difference in the
acceptance rate can be largely attributed to demographic differences in the
seniority of PIs that correlate with gender. Proposals led by senior
astronomers have a higher acceptance rate than proposals from junior
astronomers, and the fraction of senior astronomers that are women is lower
than among junior applicants. After accounting for seniority, \cite{Patat16}
found a residual systematic remained that could reflect either unaccounted
demographic differences between women and men or potentially a true gender
bias.

\citet{Lonsdale16} analyzed the results from the proposal review process for
four facilities operated in full or in part by the National Radio Astronomical
Observatory (NRAO): the Jansky Very Large Array (JVLA), the Very Long Baseline
Array (VLBA), the Green Bank Telescope (GBT) and ALMA. Analogous to the results
for HST and ESO, they found that the proposal rankings favored men over women
in ALMA Cycles 2-4, with the largest and most significant difference found in
Cycle 3. The other NRAO telescopes showed similar trends, although the
significance was lower than found for ALMA, and in some semesters, women had
higher overall rankings than men. \citet{Hunt19} extended the analysis by
\citet{Lonsdale16} to include more recent proposal rounds at the JVLA, VLBA,
and the GBT. They found that when averaged over all proposal semesters between
2012A and 2019A for the JVLA, VLBA and GBT combined, men had a statistically
significant advantage over women in the proposal scores.

The study presented here extends the analysis of the ALMA proposal rankings
conducted by \citet{Lonsdale16} in several aspects. First, the analysis is
extended to include all cycles to date (Cycles 0-6). Second, since ALMA has
a two-stage review process, the science rankings are evaluated for both the 
preliminary science assessments (Stage 1) and the final assessment resulting 
from the face-to-face review (Stage 2) to establish at which stage in the
review process any systematics are introduced. Finally, the correlation of the
proposal rankings with other variables in addition to gender, including
experience level in submitting ALMA proposals and regional affiliation of the
PIs, is investigated.

This paper is organized as follows. Section~\ref{sec:demo} describes the
salient aspects of the two-stage proposal review process adopted by ALMA and
the demographic data collected for this study. Section~\ref{sec:stage1}
explores any systematics in the proposal rankings introduced in the first stage
of the review process. Section~\ref{sec:stage2} compares the rankings between
the first and second stages of the review process to investigate any
systematics that result from the face-to-face discussion between reviewers.
Section~\ref{sec:stage2} also examines correlation between gender and the
acceptance rate of proposals into the observing queue.
Section~\ref{sec:summary} summarizes the results and briefly describes how the
ALMA review process will evolve in the near future. An analysis of the 
acceptance rate of proposals submitted by ALMA reviewers is presented in the
Appendix to investigate if the ALMA review process favors proposals submitted
by the reviewers.

\section{Proposal Review Process and Demographic Data}
\label{sec:demo}

This section describes the ALMA proposal review process and how a list of
scientific rankings is produced from the reviewer scores. The demographic data
are then described that will be used to evaluate potential systematics in the
proposal rankings against 1) the experience level of a PI in submitting ALMA
proposals, 2) the regional affiliation of the PI, and 3) the gender of the PI.

\subsection{The ALMA proposal review process}
\label{sec:review}

Similar to many other observatories, ALMA has adopted a peer-review,
panel-based system to evaluate and rank the proposals based on scientific
merit. This system has been used for seven proposal calls, starting with Cycle
0 in 2011 and continuing to Cycle 6 in 2018. Since Cycle 1, the ALMA review
panels have been split across five scientific categories: 1) Cosmology and the
high-redshift universe, 2) Galaxies and galactic nuclei, 3) Interstellar
medium, star formation, and astrochemistry, 4) Circumstellar disks and the
solar system, and 5) Stars and stellar evolution. Four categories were used in
Cycle 0, where categories 4 and 5 were combined. The number of panels in each
category has increased over the years in response to an increased number of
submitted proposals. Cycles 4-6 each contained 18 panels split across the five
categories.

Proposals are assigned to a panel by the JAO based on the science category
selected by the PI at the time of proposal submission. Further refinement in
the panel assignments can be done based on the scientific keywords selected by
the PI such that proposals with similar keywords may be grouped in a single
panel. For example, in Cycle 6, the category for Circumstellar disks and the
solar systems contained four review panels, but planetary proposals were placed
into two of the four panels. Proposals are assigned to and scored by a single
review panel. The exception is proposals for Large Programs, which are assigned
to all panels in the appropriate scientific category.

The review process proceeds in two stages. In Stage 1, panel members
review their assigned proposals and provide preliminary numerical scores on a
scale of 1 (best) to 10 (worst). Reviewers do not score proposals for which
they have a conflict of interest. Conflicts of interest can be identified by
either the JAO Proposal Handling Team (PHT) or self-declared by the reviewers.
For this study, the average Stage 1 score is computed for each
proposal\footnote{In the actual proposal review process, the JAO normalizes the
composite Stage 1 scores for each reviewer to have the same mean and standard
deviation before averaging the scores. The normalization is not done in this
study to treat the Stage 1 and Stage 2 scores in a consistent manner.}. Stage 2
of the review process consists of a face-to-face meeting of all reviewers at a
common venue. The proposals are discussed and then re-scored by each
non-conflicted reviewer. The individual scores are averaged to produce the
final Stage 2 score. 

The number of reviewers per panel has varied from six in Cycle 0, to seven in
Cycles 1-2, and to eight in Cycles 4-6. In Cycle 0, panels in Category 4 had
seven reviewers since additional expertise was needed to cover the broad range
of topics since the category also included stellar evolution. Similarly,
Category 5 had nine reviewers per panel in Cycles 5 and 6. In Cycles 0-3, each
proposal was scored by four reviewers in Stage 1. Since Cycle 4, reviewers
score all proposals in Stage 1 for which they do not have a conflict. 

Starting in Cycle 1, approximately 25-35\% of proposals that have poor scores
in the Stage 1 reviews are ``triaged" by the JAO. Unless a reviewer
``resurrects" a triaged proposal, triaged proposals are not discussed at the
face-to-face review to allow the review panels to focus their deliberations on
the better ranked proposals. The triage level is adjusted on a regional basis
in order to maintain at least a factor of two over-subscription in the
requested observing time relative to the available time for each region. In
particular, the percentage of Chilean proposals that is triaged is typically
lower than other regions since Chile has had the lowest over-subscription rate.
The non-triaged proposals are reviewed and scored by all non-conflicted
reviewers in the panel in Stage 2.

\subsection{Proposal ranks}
\label{sec:ranks}

The outcome of the review process is a merged list of proposal ranks for all
panels combined. The JAO then determines which proposals are accepted into the
observing queue using primarily the scientific rank from the review process, as
well as the time available to each region, the over-subscription in array
configurations and right ascension, and the required weather conditions needed
to carry out the observations. Proposals that are accepted into the observing
queue are assigned a priority grade (A, B, or C) while the remaining proposals
are declined. Because of the operational considerations, the priority grades do
not strictly follow the proposal rankings. Therefore, in searching for
systematics, primarily the proposal rankings are considered (see, however, the
analysis presented in Section~\ref{grades} and the Appendix). Two lists of
proposal rankings are created: i) the ranked list after the initial proposal
assessments (Stage 1), and ii) the ranked list after the face-to-face review
(Stage 2), which excludes triaged proposals. Large Program proposals,
Director's Discretionary Time proposals, and proposals submitted to the Cycle 4
Supplemental Call for the 7-m Array are excluded from the analysis since they
are reviewed in a different manner.

In each stage, the average scores from the reviewers are used to rank the
proposals within a panel between 1 (best) to $N$ (worst), where $N$ is the
number of proposals under consideration in either Stage 1 or Stage 2. The panel
rankings are then normalized by the number of proposals in the panel so that
the rankings vary between 0 (best) and 1 (worst). The rankings from the
individual panels are merged into a single list by sorting the normalized
rankings. Any ties\footnote{In generating the ranked list of proposals used to
assign priority grades, ties between proposal ranks are broken using the
Stage 2 scores, and if a tie persists, by the proposal number.} in the
normalized rankings are broken using a random number generator. The final
ranked list is then normalized from 0 (best) to 1 (worst) with steps of
1/($N$-1). A merged ranked list is created separately based on the scores in
the Stage 1 and Stage 2 process. Since triaged proposals are not re-scored by
the panels in Stage 2, these proposals are excluded from the
merged Stage 2 ranked list.

\subsection{Experience level in submitting ALMA proposals}

\citet{Patat16} found that the proposal success rate in ESO was higher for
``professional astronomers'' than for less experienced PIs (classified into
``postdocs'' and ``students''). The expertise of a PI may correlate with the
success of a proposal if established PIs are able to write a more
compelling science case based on experience or have a better understanding on
how to use ALMA optimally. On the other hand, this assumption may lead to some
element of ``prestige'' bias, where proposals led by a well-known PI are given
more favorable scores in the review process based on reputation or standing in
the community that is not based on the scientific merit of the actual proposal.

The experience factor consists of at least two components. One component is the
overall experience level of the PI, for which one measure is the year since the
PhD was obtained or the number of years as a professional astronomer. A second
component is the experience of the PI in millimeter/submillimeter
interferometry overall and with ALMA in particular, as one may expect such a PI
to understand better the capabilities of the instrument and the current state
of the field. 

While ALMA users are requested to complete a demographic profile that includes
the year of their PhD and a self-assessment of their expertise in submillimeter
astronomy and other fields, most users do not complete their profiles.
Therefore, as a surrogate for experience, the number of cycles in which a user
has {\it submitted} an ALMA proposal as PI was determined, regardless if the
proposal was ultimately accepted or not. The experience level is computed for
each user and each cycle. For example, in Cycle 6, a user with an experience
level of 1 indicates the user submitted an ALMA proposal as PI for the first
time, while a user with an experience level of 7 has submitted at least one
proposal as PI in all seven ALMA cycles and has considerable experience with
ALMA. This metric best measures the experience that a user has in submitting
ALMA proposals, but not their career standing.

The main advantage of this metric is that it can be computed in a
straightforward and consistent manner for all users and a given cycle. However,
it does not reflect the role co-investigators may have in formulating the
proposal, especially faculty advisers to students. More subtly, this experience
metric may be a biased measure in that success in one proposal cycle may
encourage additional proposals in subsequent cycles, either as positive
reinforcement or by collecting ALMA data that can be used to justify follow-up
proposals. Conversely, having a proposal declined, especially in multiple
proposal cycles, may discourage a user from submitting further proposals.

\vspace{0.1in}
\subsection{Regional affiliation}
\label{affiliation}

ALMA proposals can be submitted by anyone without regard to nationality or
affiliation. Since ALMA operations are funded by three regions (East Asia,
Europe, and North America) with cooperation of the Chilean government, there is
an inherent diversity in the ALMA user base (see Section~\ref{demo}). All PIs
self-identify their regional affiliation (Chile, East Asia, Europe, North
America, or Other) when submitting their proposals. In this context, regional
affiliation refers to the region of the host institution as opposed to the
nationality of the PI. Chilean proposals are submitted by PIs with an 
affiliation at a Chilean research institute. Proposals assigned to East Asia
consist of PIs with affiliations in Japan, Taiwan, or the Republic of Korea.
Proposals assigned to Europe consists of PIs who have affiliations in one of
the ESO member states. Proposals assigned to North America consist of PIs from
the United States, Canada, or Taiwan. Since Taiwanese agencies contributed
funding for ALMA in both East Asia and North America, Taiwan users are listed
as having a joint East Asia and North America affiliation in the proposal
process, but for the purpose of this study, they are assigned to East Asia.
Proposals from any non-ALMA regions are grouped as ``Other".

\subsection{Gender}

ALMA does not collect the gender of PIs during the proposal submission
process, although PIs can optionally enter this information as part of their
demographic profile. As mentioned previously, most PIs do not
complete their demographic profiles and therefore this information was 
gathered manually. \citet{Lonsdale16} compiled genders for ALMA PIs in Cycles
2-4 and kindly provided their database for this analysis. In collaboration with
C. Lonsdale, a small number of gender assignments were corrected, and genders
were identified for PIs from Cycles 0, 1, 5, and 6 that were not in the
database. Genders were determined by using information on the internet or
familiarity with the PI by the author or by colleagues.
Software tools to identify the gender based on the first name were also
utilized, but corroborating information was sought. While recognizing that the
subject of gender identity is complex, genders were classified as ``male'' or
``female'' for this study.

\subsection{Demographic overview}
\label{demo}

Tables~\ref{tbl:pi} and \ref{tbl:gender} summarize the regional and gender
demographics of the proposal PIs for each proposal cycle. The regional
distribution of proposals has been fairly constant throughout the first seven
cycles in that Chilean PIs have submitted \about 6\% of the proposals, East
Asian PIs \about 20\%, European PIs \about 42\%, North America \about 29\%, and
non-ALMA regions \about 3\%. The overall fraction of PIs who are women has
been increasing with time and is now nearly 34\%. Europe and especially East
Asia have seen significant increases in the percentage of female PIs from
Cycles 0 and 1. North America has seen increases in the fraction of female PIs
in the past two cycles while Chile has seen fewer female PIs recently compared
to earlier cycles.

Table~\ref{tbl:aprc} shows the regional demographics of the proposal reviewers
for each cycle. By design, reviewers are represented from all regions in
proportion to the regional shares of time (10\% for Chile, 22.5\%
for East Asia, and 33.75\% each for Europe and North America). Reviewers from
non-ALMA regions have also participated in each cycle. Relative to the
regional distribution of PIs indicated in Table~\ref{tbl:pi}, Chilean and North
America reviewers are represented in greater proportion than their share of
proposal submissions, Europe is represented less, and East Asia is about equal.
Each review panel has representation from each region and women to the extent
possible. The percentage of reviewers who are women (43\% on average) has been
consistently greater than the percentage of proposals led by women (31\% on
average). 

The JAO requests that reviewers serve 3 consecutive cycles as panel members,
which implies there is overlap in the reviewer membership from cycle to cycle.
The turnover in the reviewer membership originates from reviewers completing
three years of service, the increase in the number of reviewers in some years
to accommodate a larger number of proposals, and reviewers who choose not to
serve all three years or not in consecutive years. On average, about 45\% of
the reviewers in a given cycle did not serve in the previous cycle.

\vspace{0.6in}
\begin{deluxetable}{@{\extracolsep{4pt}}ccccccc}[h]
\tablecaption{Regional Demographics of ALMA Principal Investigators\label{tbl:pi}}
\tablehead{
   \colhead{Cycle} &
   \colhead{Number} &
   \multicolumn{5}{c}{Region}\\
 \cline{3-7}
 \colhead{} &
 \colhead{Proposals} &
 \colhead{Chile} &
 \colhead{East Asia} &
 \colhead{Europe} &
 \colhead{North America} &
 \colhead{Other}
}
\startdata
0 &  919 &  3.8\% & 19.9\% & 43.5\% & 30.5\% & 2.3\%\\
1 & 1131 &  5.7\% & 18.7\% & 43.0\% & 29.9\% & 2.7\%\\
2 & 1381 &  6.9\% & 19.7\% & 40.8\% & 30.1\% & 2.5\%\\
3 & 1578 &  7.3\% & 18.8\% & 41.6\% & 29.4\% & 2.9\%\\
4 & 1571 &  6.1\% & 21.6\% & 42.3\% & 27.1\% & 2.9\%\\
5 & 1661 &  5.3\% & 20.0\% & 42.2\% & 29.6\% & 2.9\%\\
6 & 1836 &  5.8\% & 20.0\% & 42.6\% & 28.5\% & 3.1\%\\
\enddata
\tablecomments{Table shows the percentage of proposal PIs from each region.}
\end{deluxetable}

\begin{deluxetable}{@{\extracolsep{4pt}}ccccccccccccc}[h]
\tablecaption{Gender Demographics of ALMA Principal Investigators\label{tbl:gender}}
\tablehead{
   \colhead{Cycle} &
   \multicolumn{1}{c}{Chile} &
   \multicolumn{1}{c}{East Asia} &
   \multicolumn{1}{c}{Europe} &
   \multicolumn{1}{c}{North America} &
   \multicolumn{1}{c}{Other} &
   \multicolumn{1}{c}{All}
}
\startdata
  0 &  28.6\% & 16.9\% & 30.5\% & 32.1\% & 23.8\% & 28.1\% \\
  1 &  24.6\% & 14.6\% & 30.2\% & 32.2\% & 16.7\% & 27.2\% \\
  2 &  25.3\% & 24.6\% & 35.7\% & 33.7\% & 17.1\% & 31.7\% \\
  3 &  14.8\% & 26.3\% & 36.2\% & 32.3\% & 33.3\% & 31.6\% \\
  4 &  19.8\% & 24.7\% & 36.7\% & 30.5\% & 33.3\% & 31.4\% \\
  5 &  20.5\% & 25.2\% & 36.9\% & 35.8\% & 22.9\% & 33.0\% \\
  6 &  19.8\% & 26.6\% & 36.8\% & 37.5\% & 23.2\% & 33.6\% \\
\enddata
\tablecomments{Table lists the percentage of PIs who are women in each region.}
\end{deluxetable}

\begin{deluxetable}{@{\extracolsep{4pt}}ccccccccc}[h]
\tabletypesize{\normalsize}
\tablecaption{Demographics of the ALMA Proposal Reviewers\label{tbl:aprc}}
\tablehead{
   \colhead{Cycle} &
   \colhead{Number} &
   \multicolumn{5}{c}{Region} &
   \multicolumn{2}{c}{Gender}\\
 \cline{3-7}
 \cline{8-9}
 \colhead{} &
 \colhead{Reviewers} &
 \colhead{Chile} &
 \colhead{East Asia} &
 \colhead{Europe} &
 \colhead{North America} &
 \colhead{Other} &
 \colhead{Female} &
 \colhead{Male}
}
\startdata
0 &  49 & 10.2\% & 20.4\% & 36.7\% & 28.6\% & 4.1\% & 40.8\% & 59.2\% \\
1 &  77 & 10.4\% & 22.1\% & 32.5\% & 32.5\% & 2.6\% & 39.0\% & 61.0\% \\
2 &  77 & 10.4\% & 22.1\% & 33.8\% & 32.5\% & 1.3\% & 40.3\% & 59.7\% \\
3 &  96 & 10.4\% & 21.9\% & 33.3\% & 33.3\% & 1.0\% & 44.8\% & 55.2\% \\
4 & 145 &  9.7\% & 21.4\% & 33.1\% & 33.1\% & 2.8\% & 47.6\% & 52.4\% \\
5 & 146 &  9.6\% & 19.9\% & 35.6\% & 31.5\% & 3.4\% & 47.9\% & 52.1\% \\
6 & 146 & 10.3\% & 22.6\% & 32.9\% & 29.5\% & 4.8\% & 41.8\% & 58.2\% \\
\enddata
\tablecomments{Table shows the regional and gender distribution of the ALMA 
reviewers.}
\end{deluxetable}

\section{Analysis of the Stage 1 Rankings}
\label{sec:stage1}

This section analyzes the Stage 1 proposal rankings to identify any systematics
based on experience level (Section~\ref{analysis:experience}), regional
affiliation (Section~\ref{analysis:region}), and gender
(Section~\ref{analysis:gender}) that are introduced in the preliminary reviewer
scores. 
Potential systematics are examined by analyzing the cumulative distribution of
proposal ranks; e.g., comparing the cumulative distribution of proposal ranks
for female and male PIs. This approach has the advantage that differences
anywhere along the cumulative profiles can be captured. The number of
cumulative distributions being compared can be as few as two when comparing by
gender, to as many as five for regional comparisons, and seven for
experience-level comparisons.

The Anderson-Darling $k$-sample test \citep{Scholz87} as implemented in {\tt
scipy} was used to measure the difference between cumulative distributions. The
Anderson-Darling test statistic was then used to compute the probability (\pad,
$0 \le p_{AD} \le 1$)
that the $k$ samples are drawn from the same (but unspecified) population using
the {\tt pval} function within the {\tt kSamples} package designed for {\tt R}.
A low value of \pad\ suggests that the $k$ samples are drawn from different
distributions while a high value of \pad\ suggests that the $k$ samples have
similar distributions. Any differences in the cumulative ranks are arbitrarily
defined as ``significant" if the probability that the distributions are drawn
from the same population is \pad$<0.01$ and ``marginally significant" if the
probability is $0.01\le$\pad$\le0.10$.

\subsection{Experience level}
\label{analysis:experience}

\begin{figure}[h]
\epsscale{1.1}
\plotone{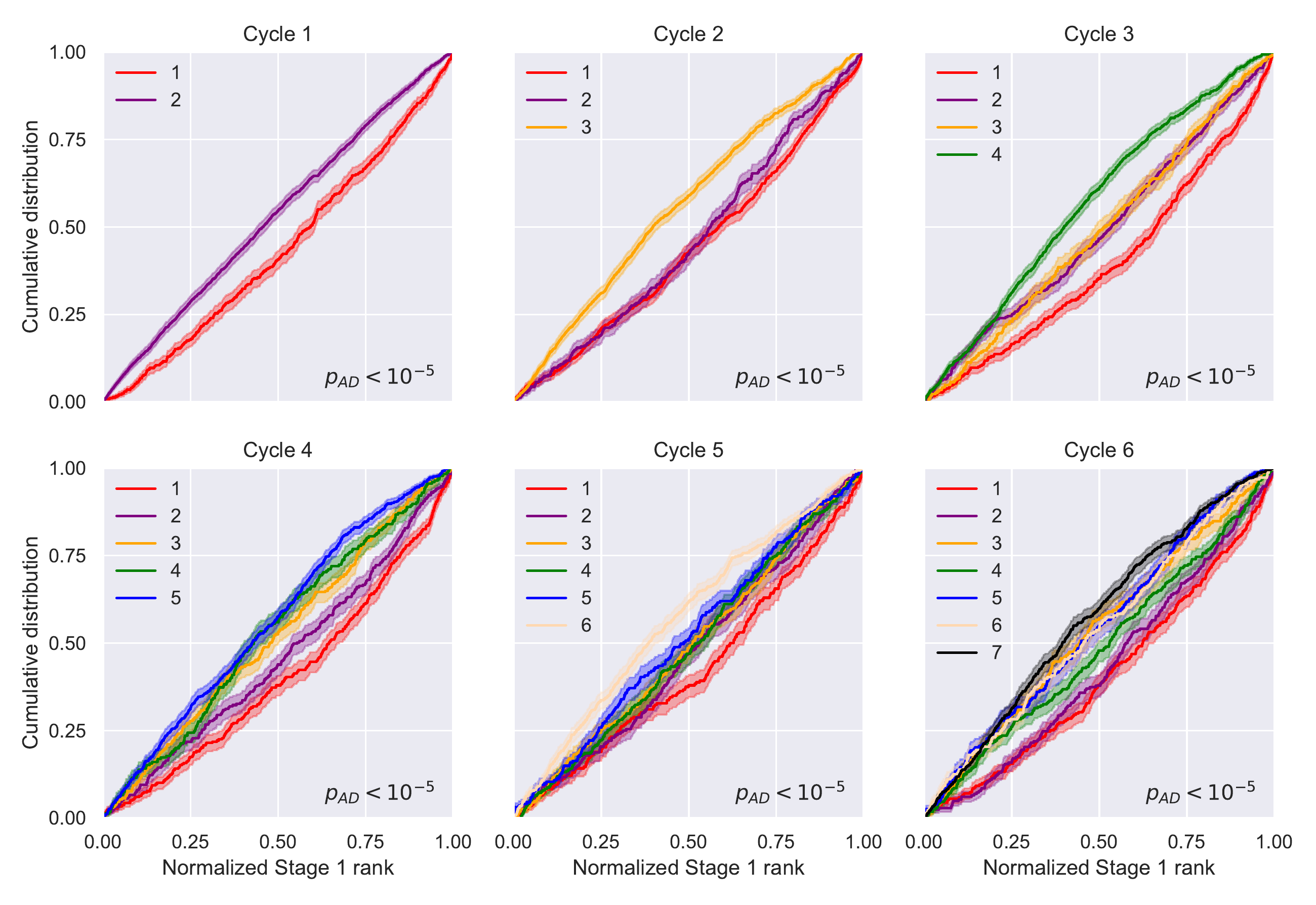}
\caption{
  \label{fig:ad_experience}
  Normalized cumulative distribution of Stage 1 proposal ranks by experience
  level for each cycle. The normalized ranks vary between 0 (best) to 1 
  (worst). The shaded region indicates the 68.3\% confidence interval computed 
  using the beta function. The probability from the Anderson-Darling 
  $k$-sample test that the distributions within a cycle are drawn from the 
  same population is indicated in the lower right corner of each panel.
} 
\end{figure}

Figure~\ref{fig:ad_experience} shows the cumulative distribution of Stage 1
proposal ranks by experience level for Cycles 1-6, where the experience level
is at the time of the indicated cycle. Cycle 0 is not shown since all PIs
submitted proposals for the first time. In this figure and all similar figures
that follow, the solid line shows the cumulative distribution of ranks and the
shaded region shows the 68.3\% confidence interval (i.e., ``1$\sigma$'')
computed using the beta function. Since the best-ranked proposals have a
normalized rank of 0 and the poorest-ranked proposals have a normalized rank of
1, curves shifted to the upper left have better overall ranks compared to
curves shifted to the lower right. The probability (\pad) that the curves are
drawn from the same population is indicated in the lower right of each panel. 

As an example, the upper left panel in Figure~\ref{fig:ad_experience} shows
that in Cycle 1, PIs who submitted proposals in both Cycles 0 and 1
(experience level = 2) had better proposal ranks than first-time PIs in Cycle 1
(experience level = 1). The trend is present in each of the first, second, and
third quartiles of the cumulative distributions. The difference in proposal
ranks is significant in that the probability that the two distributions are
drawn from the same population is \pad$<10^{-5}$. Each subsequent cycle shows
the same trend in that PIs who have submitted proposals in more cycles tend to
have better proposal ranks than PIs who have submitted proposals in fewer
proposal cycles. The strongest and most persistent trend is that first-time PIs
have the poorest proposal ranks, while PIs who submit proposals every cycle
have the best proposal ranks. Proposal ranks for intermediate experience levels
are also generally correlated with experience level. While not shown here,
these basic trends are typically present within each region separately,
although there are singular cycles where the trends are not strictly
followed within a region.

\clearpage
\subsection{Regional affiliation}
\label{analysis:region}

\begin{figure}[h]
\epsscale{1.19}
\plotone{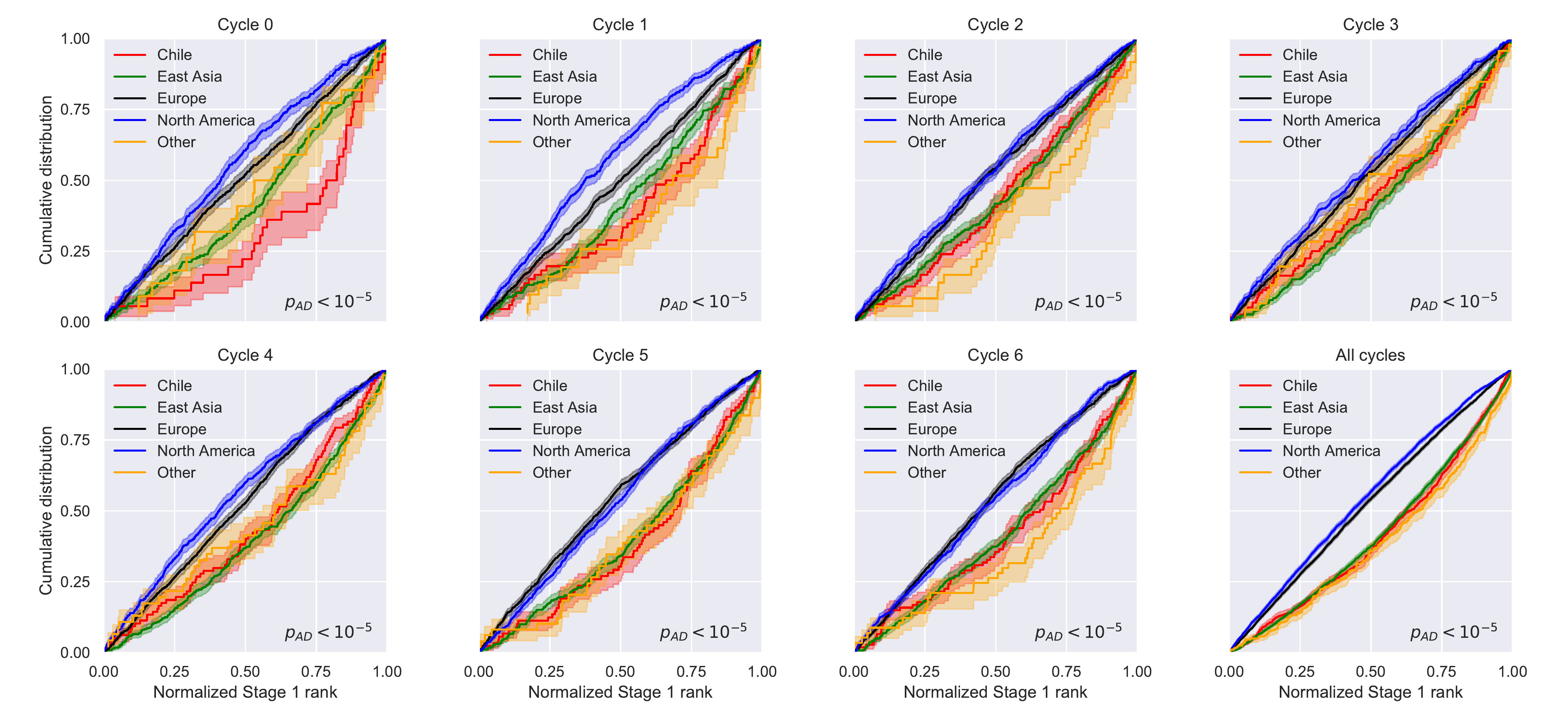}
\caption{
  \label{fig:ad_executive}
  Normalized cumulative distribution of Stage 1 proposal ranks by
  regional affiliation for each cycle. 
}
\end{figure}

Figure~\ref{fig:ad_executive} shows the cumulative distribution of Stage 1
proposal ranks by regional affiliation for Cycles 0--6. Each cycle exhibits the
same trend in that PIs from North America and Europe have better proposal
rankings overall than PIs from Chile, East Asia, and other regions. The trend
is present and significant in each cycle. The differences appear to
moderate somewhat in Cycles 2 and 3 but increase in Cycles 4-6.

In Cycles 0 and 1, North American PIs had better ranked proposals
than European PIs. However, in later cycles, the differences diminished. 
Averaged over all cycles, there is a marginal tendency for North American 
proposals to have better ranks than European proposals, but the tendency
vanishes if Cycles 0 and 1 are excluded. No significant difference in the 
proposals ranks are observed for Chilean and East Asian PIs within a
cycle or when averaged over all cycles (\pad=0.56).

\begin{figure}[h]
\epsscale{1.19}
\plotone{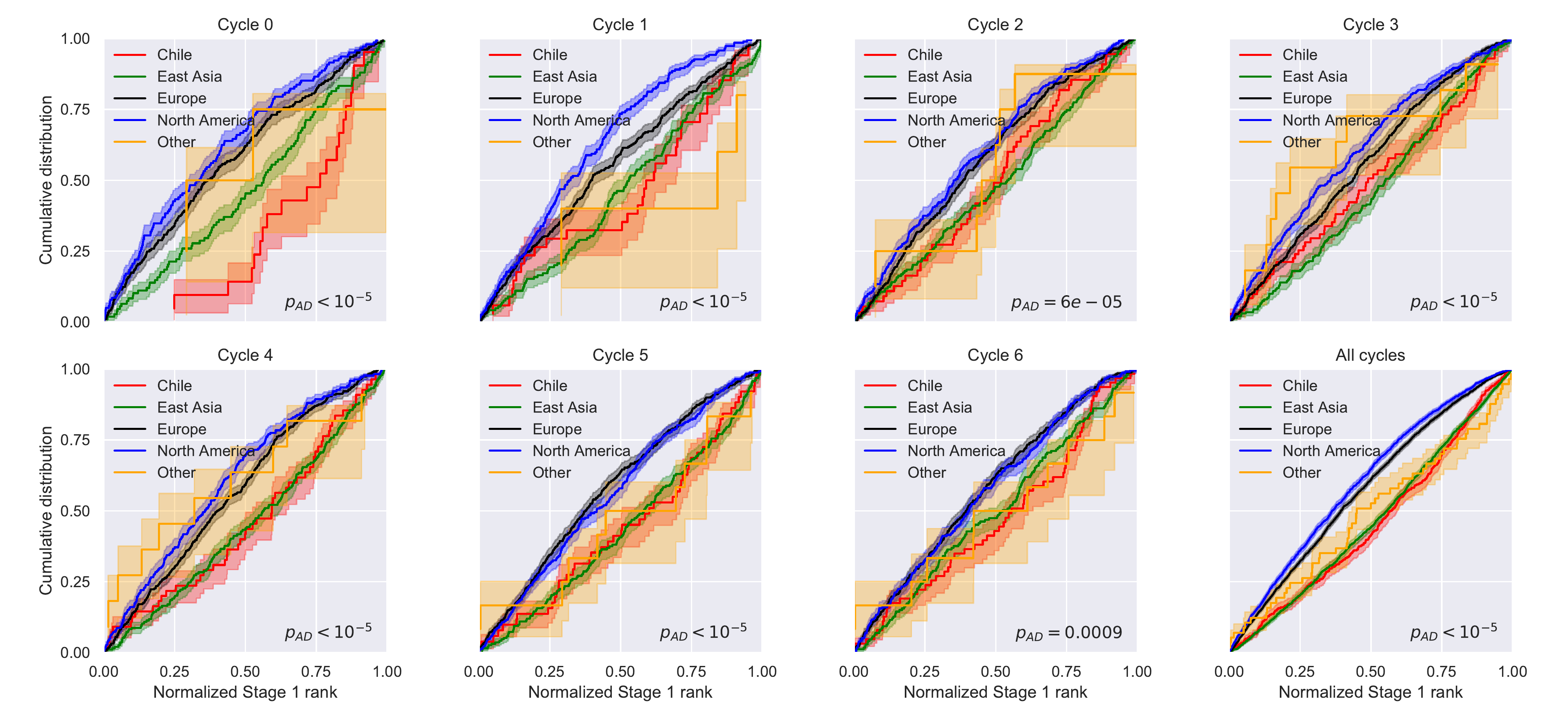}
\caption{
  \label{fig:ad_executive_567}
  Normalized cumulative distribution of Stage 1 proposal ranks by
  regional affiliation for each cycle for PIs who have submitted
  proposals in 5 or more cycles, which represents the most experienced
  ALMA users.
}
\end{figure}

\begin{figure}[h]
\epsscale{1.19}
\plotone{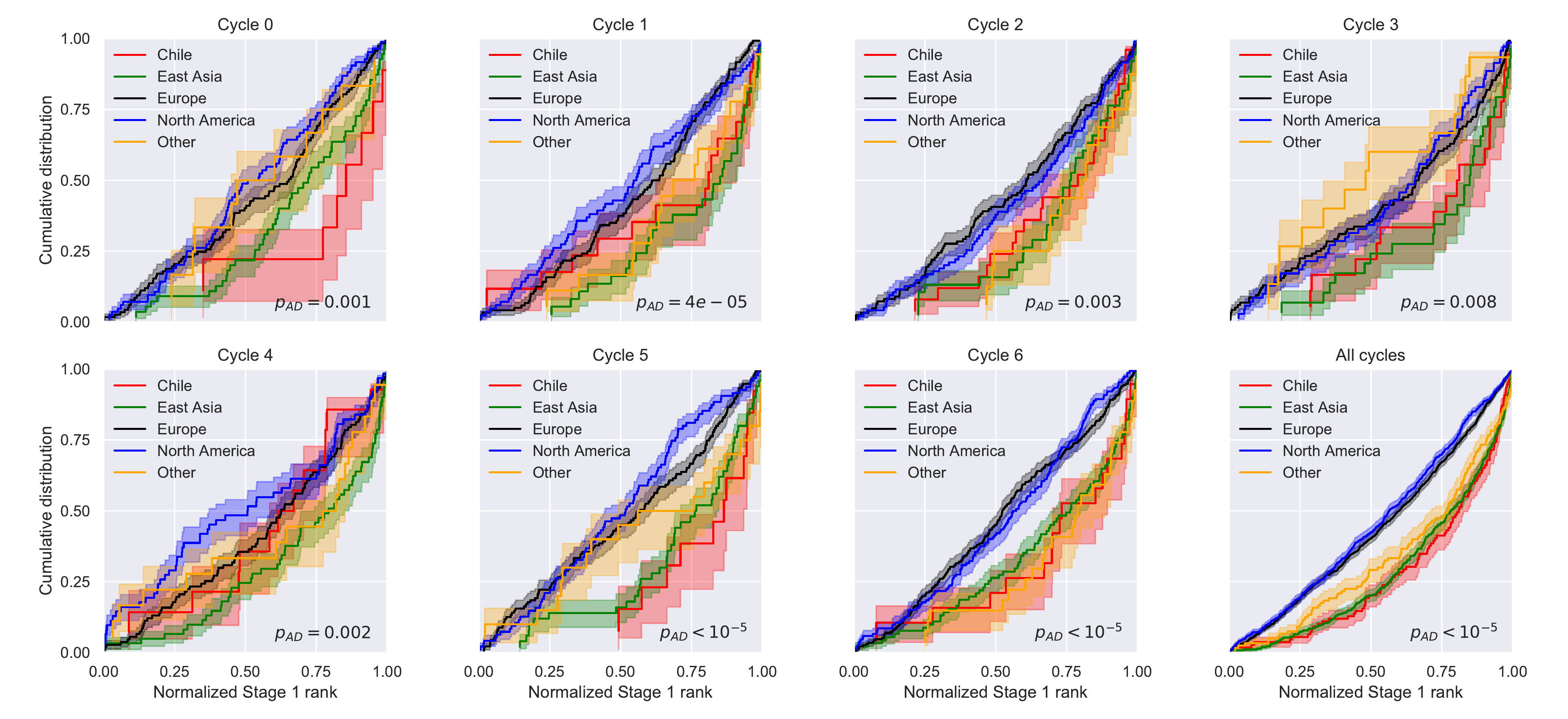}
\caption{
  \label{fig:ad_executive_12}
  Normalized cumulative distribution of Stage 1 proposal ranks by
  regional affiliation for each cycle for PIs who have submitted
  proposals in only 1 or 2 cycles, which represents the least experienced
  ALMA users.
}
\end{figure}

Differences in the relative proposal ranks by region transcend across the
experience levels of the PIs. Figures~\ref{fig:ad_executive_567} shows the
cumulative proposal ranks by region for the most experienced PIs, defined as
users who have submitted proposals in at least five of the seven cycles. 
Similarly, Figure~\ref{fig:ad_executive_12} shows the results for PIs
who have submitted proposals in only one or two cycles to select inexperienced
ALMA users. In both subsamples, PIs from Europe and North America have 
significantly better proposal ranks than PIs from other regions.

\clearpage

\subsection{Gender}
\label{analysis:gender}

\begin{figure}[h]
\epsscale{1.19}
\plotone{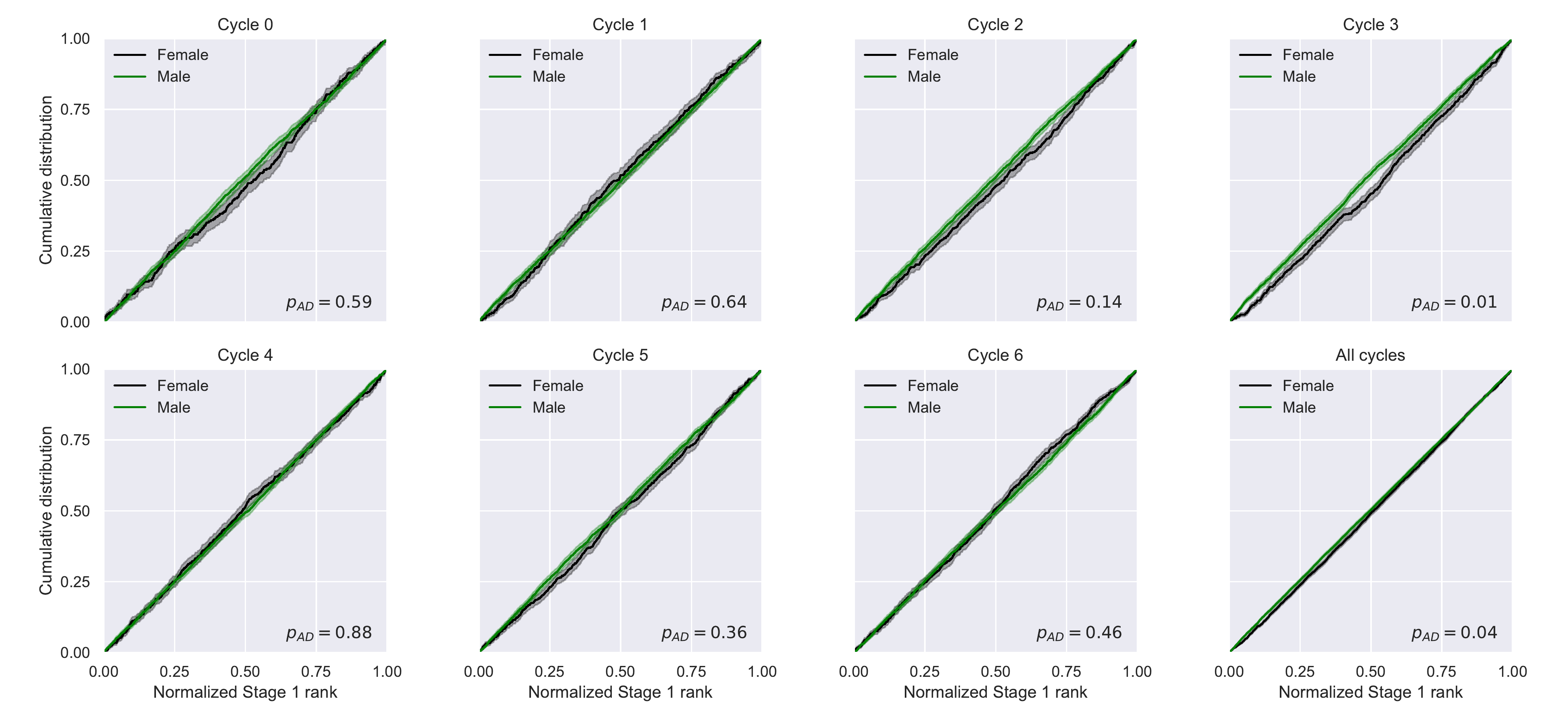}
\caption{
  \label{fig:ad_gender}
  Normalized cumulative distribution of Stage 1 proposal ranks by gender 
  for each cycle.
}
\end{figure}

Figure~\ref{fig:ad_gender} shows the cumulative distribution of Stage 1
proposal ranks by gender for each cycle. No significant difference between the
proposal ranks for women or men exists in any individual cycle. Consistent with
\citet{Lonsdale16}\footnote{The \citet{Lonsdale16} results differ in detail 
compared to this paper since they used a ranked list of proposals based on the 
Stage 2 results merged with the triaged proposals. This paper analyzes both 
the Stage 1 and Stage 2 results, but does not merge triaged proposals with the
Stage 2 rankings.}, proposals led by men had better ranks than proposals led by
women in Cycle 3 with marginal significance (\pad=0.01) and to a lesser extent
in Cycle 2 (\pad=0.14). Averaged over all cycles, the probability that the
distribution of proposal ranks are different based on gender is \pad=0.04 and
is marginally significant. 

The results in Cycle 3 stand out in that men had better ranks than women when
measured at the first, second, and third quartile points in the cumulative
rankings. It is unclear why Cycle 3 would be noteworthy in this regard.
No fundamental change was introduced in the review process itself, and the
percentage of proposals from women and the percentage of women reviewers were
in line with other cycles. 

While the results of the \citet{Lonsdale16} paper were posted after the Cycle 4
proposal review, preliminary results had been presented to the community 
before the Cycle 4 proposal review and the JAO communicated the results to the
Cycle 4 reviewers at the Stage 2 orientation meeting. The Cycle 5 and 6
reviewers received guidance on the results and the role of unconscious bias in
the written Stage 1 review instructions and at the Stage 2 orientation meeting.
Once the presence of systematics began to be communicated (in Cycles 4-6), no
discernible differences in the Stage 1 proposal rankings between women and men
are evident in any individual cycle or when the three cycles are combined
(\pad=0.73). However, it is unclear that alerting the community after Cycle 3
actually contributed to reducing the gender-based systematic or if the results
from Cycle 3 were just a statistical outlier.

One difficulty in interpreting Figure~\ref{fig:ad_gender} is that any
systematics between genders are much smaller than the systematics present
by experience level and regional affiliation. Thus, changes in the underlying
experience or regional demographics of the PIs can be responsible for the 
difference in the proposal ranks by gender \citep[see also][]{Patat16}. Given
these considerations, subsets of the data are analyzed to further examine
possible systematics in order to isolate the impact of gender alone.

\begin{figure}
\epsscale{1.19}
\plotone{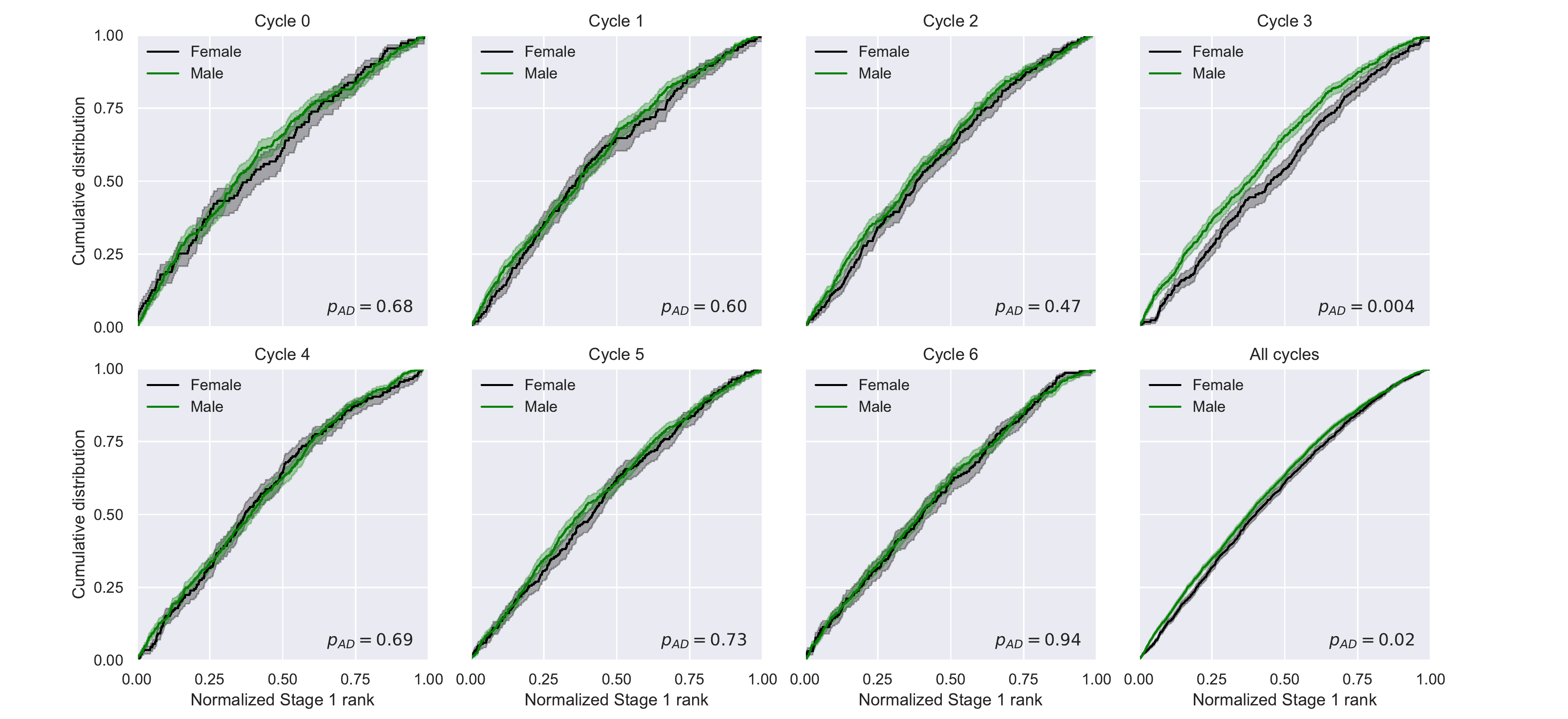}
\caption{
  \label{fig:ad_eu_na}
  Normalized cumulative distribution of Stage 1 proposal ranks for women and 
  men in Europe and North America for Cycles 0-6 and all cycles combined.
  The results are shown only for PIs who have submitted an ALMA proposal in 
  at least 5 cycles.
}
\end{figure}

Figure~\ref{fig:ad_eu_na} shows the cumulative distribution of proposal ranks
for PIs from Europe and North America who have submitted proposals in at least
five cycles. Europe and North America were grouped together since they share
similar proposal ranks overall. The most experienced PIs were selected since
the fraction of women PIs has been increasing over time and first-time PIs
typically have poorer proposal ranks. Figure~\ref{fig:ad_eu_na}
show that even among experienced PIs, women have had poorer
ranked proposals than men when averaged over all cycles, but with marginal
significance. The difference is driven by the significant difference found in
Cycle 3. If Cycle 3 is excluded, any differences in the proposal rankings
between experience female and male PIs are insignificant even when averaged 
over the other cycles (\pad=0.25).

\begin{figure}
\epsscale{1.19}
\plotone{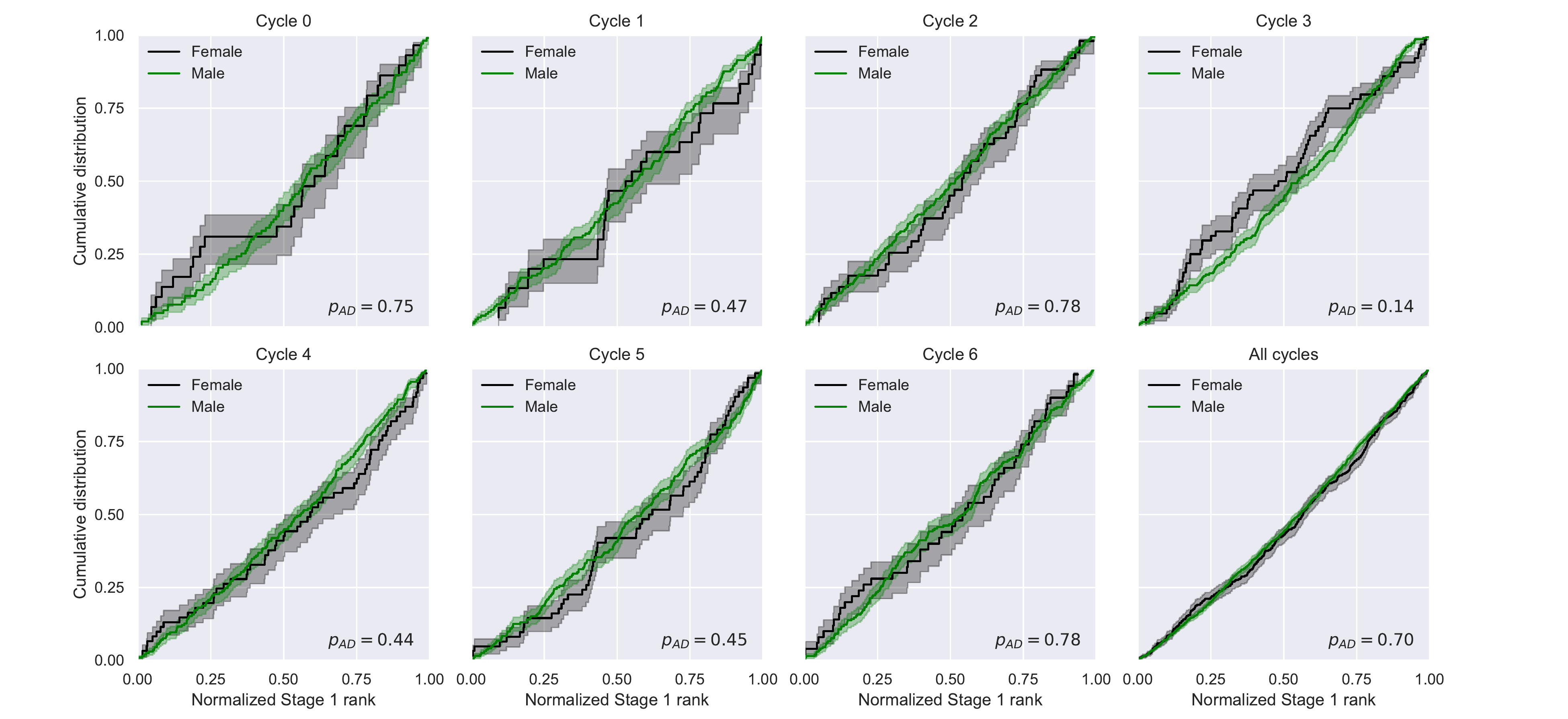}
\caption{
  \label{fig:ad_cl_ea_other}
  Normalized cumulative distribution of Stage 1 proposal ranks for women and 
  men in Chile, East Asia, and non-ALMA regions for Cycles 0-6 and all cycles combined.
  The results are shown only for PIs who have submitted an ALMA proposal in 
  at least 5 cycles.
}
\end{figure}

Figure~\ref{fig:ad_cl_ea_other} shows the difference in the proposal ranks for
experienced PIs from Chile, East Asia, and non-ALMA regions. No significant
difference in the proposal rankings are found even if averaged over all cycles
(\pad=0.70). Interestingly, in Cycle 3, women from Chile, East Asia, and
non-ALMA regions had better proposal ranks than men, although the difference is
not significant. Nonetheless, this is opposite of the trend found amount
European and North American PIs. These different results suggests that any
potential biases are complex and cannot be simply cast by gender alone.

\begin{figure}[h]
\epsscale{1.19}
\plotone{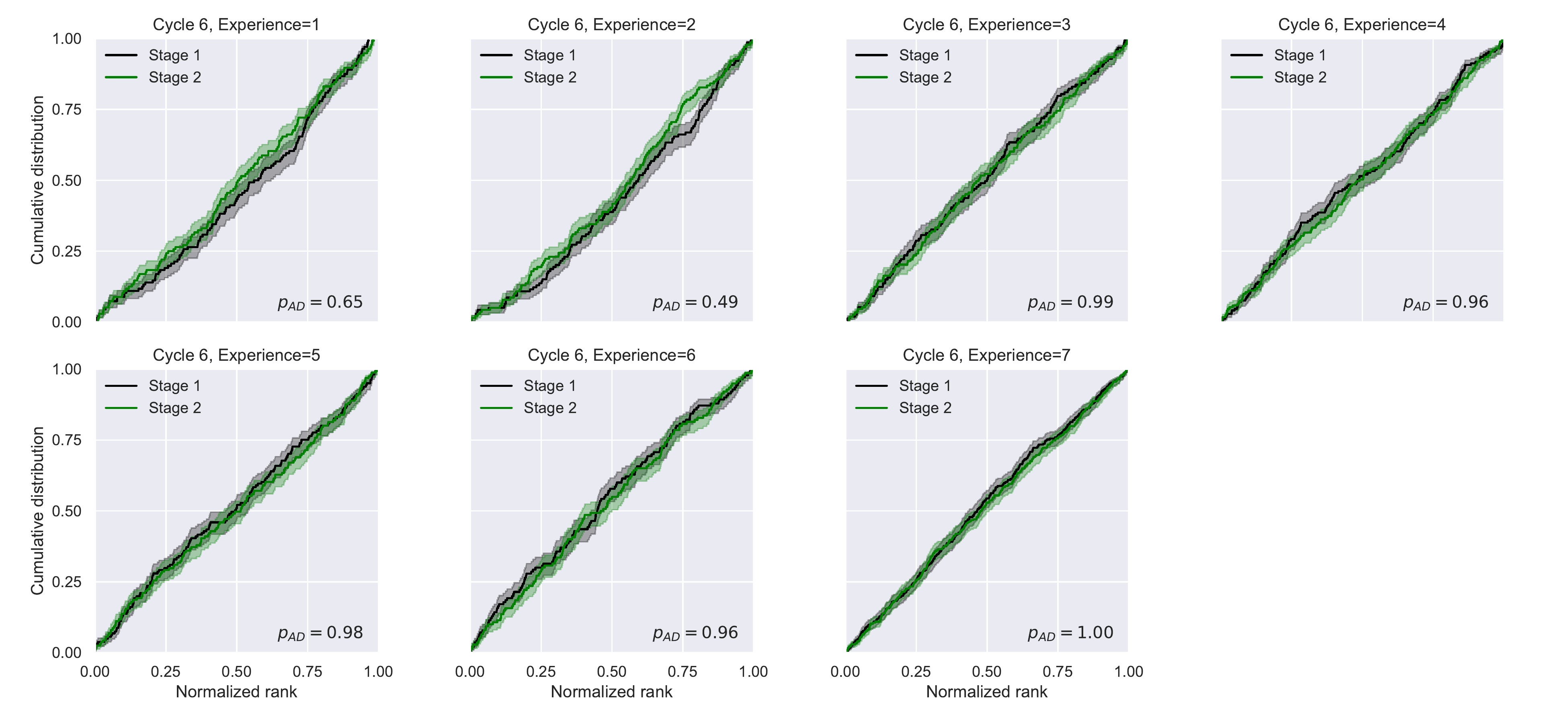}
\caption{
  \label{fig:ad_experience_nontriaged}
  Normalized cumulative distribution of Stage 1 and Stage 2 proposal ranks 
  (solid curves) in Cycle 6 for different experience levels. Only non-triaged 
  proposals are shown. 
}
\end{figure}

\clearpage
\section{Analysis of the Stage 2 Results}
\label{sec:stage2}

The analysis in Section~\ref{sec:stage1} showed that systematics are introduced
in the Stage 1 rankings, especially with respect to experience level and
regional affiliation. In this section, any systematics in the Stage 2 results
are analyzed. First, the triage proposals are analyzed to assess if any
systematics are present with respect to gender (Section~\ref{triaged}). Then
the Stage 1 and Stage 2 rankings for the non-triaged proposals are compared to
determine if any of the systematics identified in the Stage 1 rankings are
amplified or reduced as a result of the face-to-face discussion
(Section~\ref{stage1_2}). Finally, the proposals added to the observing queue
are analyzed to determine the acceptance rate of proposals with respect
to gender (Section~\ref{grades}).

\subsection{Triaged proposals}
\label{triaged}

As described in Section~\ref{sec:review}, poorly ranked proposals are triaged
after the Stage 1 review to reduce the number of proposals discussed at Stage
2. The JAO identifies the proposals that are triaged, but upon request the
reviewers may resurrect a triaged proposal and have it discussed in the
face-to-face review. Table~\ref{tbl:triage} lists the fraction of triaged
proposals that have a female PI. This triage fraction by gender should not be
compared directly to the overall fraction of proposals led by women (see
Table~\ref{tbl:gender}) to identify potential biases since the demographics of
triaged proposals are not in general the same as the overall proposals.
This is primarily because the gender balance differs between regions, and
the fraction of proposals triaged per region will differ.

To account for the demographics of the triaged proposals, the expected number 
of triaged proposals with female PIs was estimated as the number of triaged 
proposals in a given demographic group multiplied by the fraction of all 
proposals with a female PI in that group. Formally, the expected fraction of 
triaged proposals with female PIs ($f_{t,expected}$) is
\begin{equation}
f_{t,expected} = 
\frac{\sum\limits_{Region}\sum\limits_{Experience}\sum\limits_{Category} f(R,E,C) N_{triage}(R,E,C)}
{\sum\limits_{Region}\sum\limits_{Experience}\sum\limits_{Category} N_{triage}(R,E,C)},
\end{equation}
\noindent
where $f(R,E,C)$ is the fraction of PIs with regional affiliation $R$,
experience level $E$, and science category $C$ that are female;
$N_{triage}(R,E,C)$ is the total number of triaged proposals in the demographic
group, excluding proposals resurrected by reviewers. The uncertainties in the
expected fraction were estimated assuming Poisson statistics on the number of
female PIs used to compute $f(R,E,C)$.

The expected fraction of triaged proposals for female PIs is listed in of
Table~\ref{tbl:triage} by region and for all regions combined. In Cycles 1-5,
female PIs across all regions had a larger fraction of the triaged proposals
than expected based on the demographics and share of the proposals. The
difference was largest in Cycle 3 as could have been anticipated based on the
Stage 1 rankings (see Figure~\ref{fig:ad_gender}). Nonetheless, the differences
are not statistically significant in any given cycle. Only in Cycle 6 did
female PIs have a lower percentage of the triaged proposals than expected based
on the model. These basic trends are seen in East Asia, Europe, and North
America, although there are individual cycles where female PIs in East Asia
(Cycles 1 and 2) and North America (Cycles 5 and 6) had fewer proposals triaged
than expected. Notably in Europe, female PIs had a greater fraction of
proposals triaged than expected in each cycle. The number of triaged proposals
in Chile and non-ALMA regions are too small to identify any meaningful trends.

\begin{deluxetable}{@{\extracolsep{1pt}}ccccccccccccc}
\tabletypesize{\scriptsize}
\tablecaption{Gender Demographics of Triaged Proposals\label{tbl:triage}}
\tablehead{
   \colhead{Cycle} &
   \multicolumn{2}{c}{Chile} &
   \multicolumn{2}{c}{East Asia} &
   \multicolumn{2}{c}{Europe} &
   \multicolumn{2}{c}{North America} &
   \multicolumn{2}{c}{Other} &
   \multicolumn{2}{c}{All regions}\\
   \cline{2-3}
   \cline{4-5}
   \cline{6-7}
   \cline{8-9}
   \cline{10-11}
   \cline{12-13}
   \colhead{} &
   \colhead{$f_{t}$} &
   \colhead{$f_{t,expected}$} &
   \colhead{$f_{t}$} &
   \colhead{$f_{t,expected}$} &
   \colhead{$f_{t}$} &
   \colhead{$f_{t,expected}$} &
   \colhead{$f_{t}$} &
   \colhead{$f_{t,expected}$} &
   \colhead{$f_{t}$} &
   \colhead{$f_{t,expected}$} &
   \colhead{$f_{t}$} &
   \colhead{$f_{t,expected}$}
}
\startdata
   1  & 32.1\% & 21.4\% $\pm$ 6.2\% & 12.9\% & 15.0\% $\pm$ 2.8\% & 34.5\% & 30.3\% $\pm$ 2.6\% & 30.9\% & 33.3\% $\pm$ 3.5\% &  8.3\% & 10.0\% $\pm$ 7.1\% & 27.8\%  & 26.0\% $\pm$ 1.6\% \\
   2  & 44.4\% & 38.9\% $\pm$ 8.7\% & 23.3\% & 25.8\% $\pm$ 3.3\% & 38.5\% & 37.1\% $\pm$ 2.8\% & 41.9\% & 34.8\% $\pm$ 3.2\% & 27.8\% & 22.2\% $\pm$ 9.4\% & 35.2\%  & 32.9\% $\pm$ 1.7\% \\
   3  &  0.0\% & 26.7\% $\pm$ 12.5\% & 27.6\% & 23.2\% $\pm$ 3.8\% & 41.5\% & 36.7\% $\pm$ 2.5\% & 41.5\% & 32.9\% $\pm$ 3.0\% & 18.8\% & 26.5\% $\pm$ 8.0\% & 39.3\%  & 34.2\% $\pm$ 1.8\% \\
   4  &  0.0\% & 20.0\% $\pm$ 20.0\% & 29.8\% & 25.5\% $\pm$ 3.1\% & 39.4\% & 36.7\% $\pm$ 2.6\% & 32.7\% & 31.6\% $\pm$ 3.2\% & 29.4\% & 32.6\% $\pm$ 12.1\% & 34.5\%  & 32.0\% $\pm$ 1.7\% \\
   5  & \nodata &    \nodata & 32.8\% & 27.8\% $\pm$ 3.3\% & 43.2\% & 38.0\% $\pm$ 2.6\% & 35.4\% & 35.8\% $\pm$ 3.0\% & 25.0\% & 22.4\% $\pm$ 7.7\% & 37.1\%  & 33.5\% $\pm$ 1.7\% \\
   6  & 22.6\% & 24.6\% $\pm$ 6.8\% & 31.0\% & 28.3\% $\pm$ 3.1\% & 38.0\% & 37.3\% $\pm$ 2.4\% & 32.4\% & 38.1\% $\pm$ 3.0\% & 14.7\% & 16.2\% $\pm$ 5.7\% & 32.7\%  & 33.4\% $\pm$ 1.5\% \\
\enddata
\tablecomments{The table lists the fraction of triaged proposals with a female PI ($f_{t}$) 
and the expected fraction ($f_{t,expected}$) given the demographics of the triaged proposals, 
as described in the text. The results are given for each region and all regions combined. 
Cycle 0 is not listed since no proposals were triaged in that cycle.
}
\end{deluxetable}

\subsection{Comparison of Stage 1 and Stage 2 rankings}
\label{stage1_2}

This section compares the Stage 1 rankings with the Stage 2 rankings to
determine if the systematics identified in Stage 1 change significantly as a
result of the face-to-face discussion. The impact of the face-to-face
discussions was assessed by comparing the cumulative Stage 1 and Stage 2
proposal rankings of the non-triaged proposals. The Stage 1 proposal rankings
for the non-triaged proposals were extracted and then renormalized on a scale
of 0 to 1 (see Section~\ref{sec:ranks}). This renormalization was needed to
eliminate systematic differences between the Stage 1 and Stage 2 rankings since
triaged proposals have preferentially poorer ranks by design.

Figure~\ref{fig:ad_experience_nontriaged} shows the cumulative distributions of
Cycle~6 proposal rankings in Stage 1 and Stage 2 grouped by experience level.
For each experience level, the cumulative distributions for the Stage 1 and
Stage 2 normalized ranks are similar and none of the differences are considered
even marginally significant. While not shown here, previous cycles have similar
results. The ranks for individual proposals did in fact change between the
Stage 1 and Stage 2 reviews, but Figure~\ref{fig:ad_experience_nontriaged}
indicates no systematic differences were introduced based on the experience
level of the PI.

\begin{figure}[h]
\epsscale{1.19}
\plotone{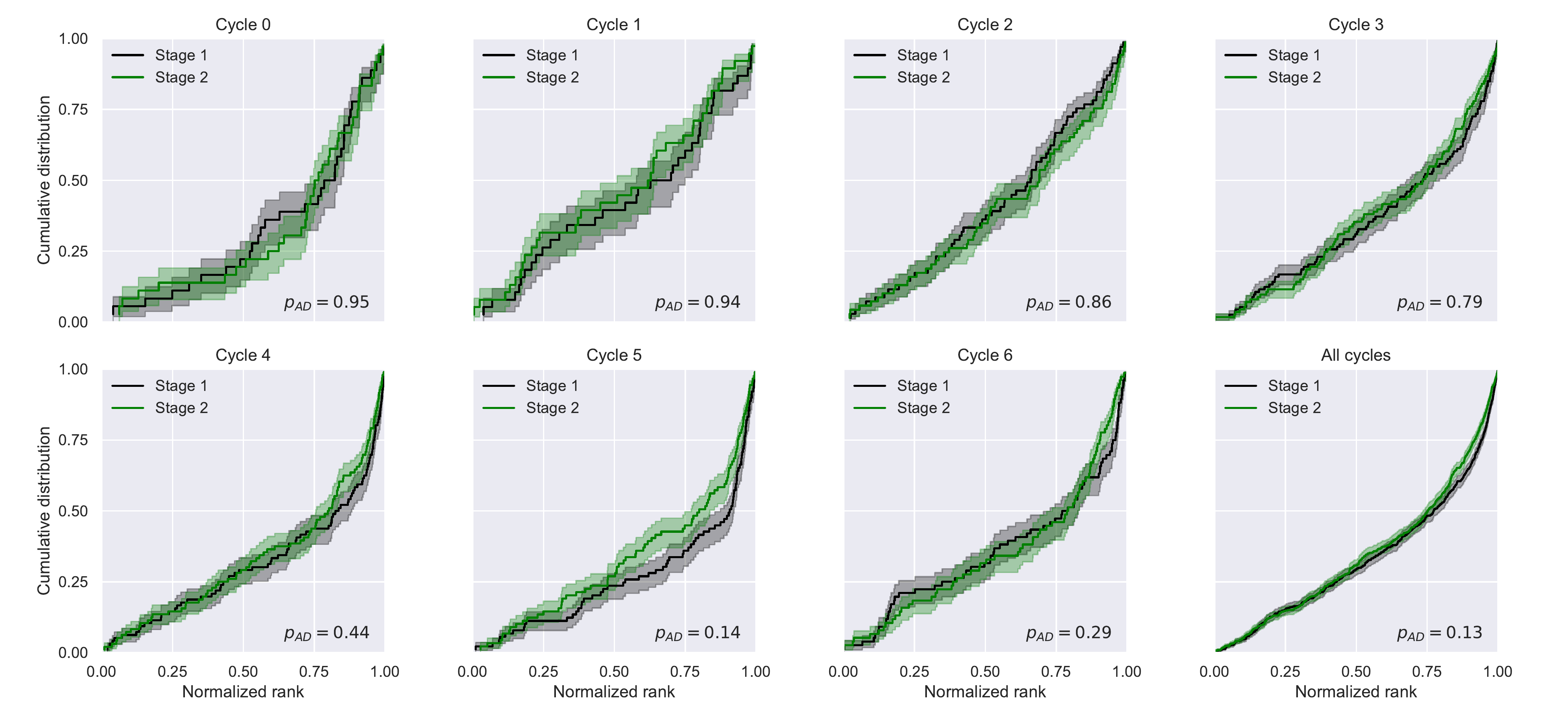}
\caption{
  \label{fig:ad_cl_nontriaged}
  Cumulative distribution of Stage 1 and Stage 2 proposal ranks for 
  non-triaged proposals led by a Chilean PI in Cycles 0--6 and for all
  cycles combined. 
}
\end{figure}

\begin{figure}[h]
\epsscale{1.19}
\plotone{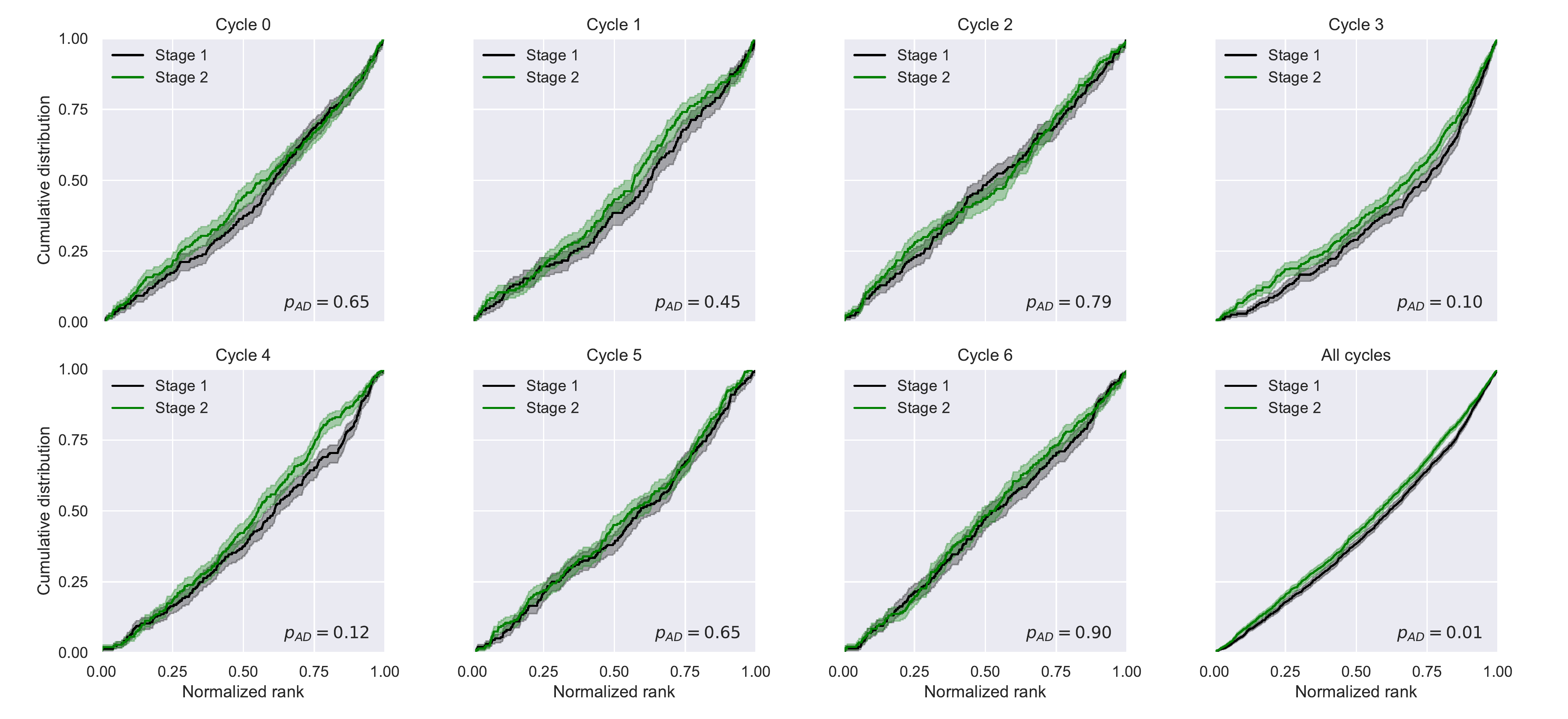}
\caption{
  \label{fig:ad_ea_nontriaged}
  Cumulative distribution of Stage 1 and Stage 2 proposal ranks for non-triaged proposals 
  led by an East Asian PI in Cycles 0--6 and for all cycles combined.
}
\end{figure}

\begin{figure}[h]
\epsscale{1.19}
\plotone{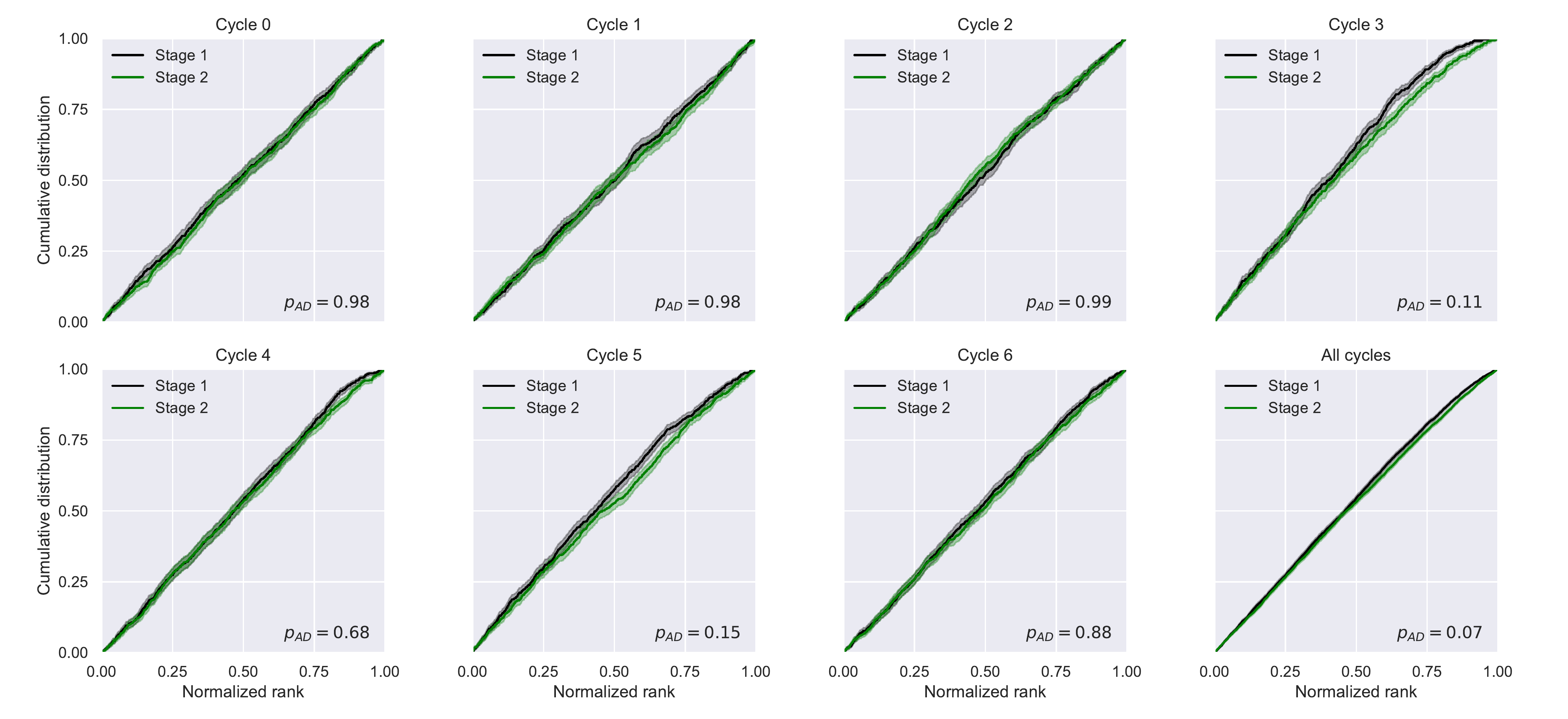}
\caption{
  \label{fig:ad_eu_nontriaged}
  Cumulative distribution of Stage 1 and Stage 2 proposal ranks for non-triaged proposals 
  led by an European PI in Cycles 0--6 and for all cycles combined.
}
\end{figure}

\begin{figure}[h]
\epsscale{1.19}
\plotone{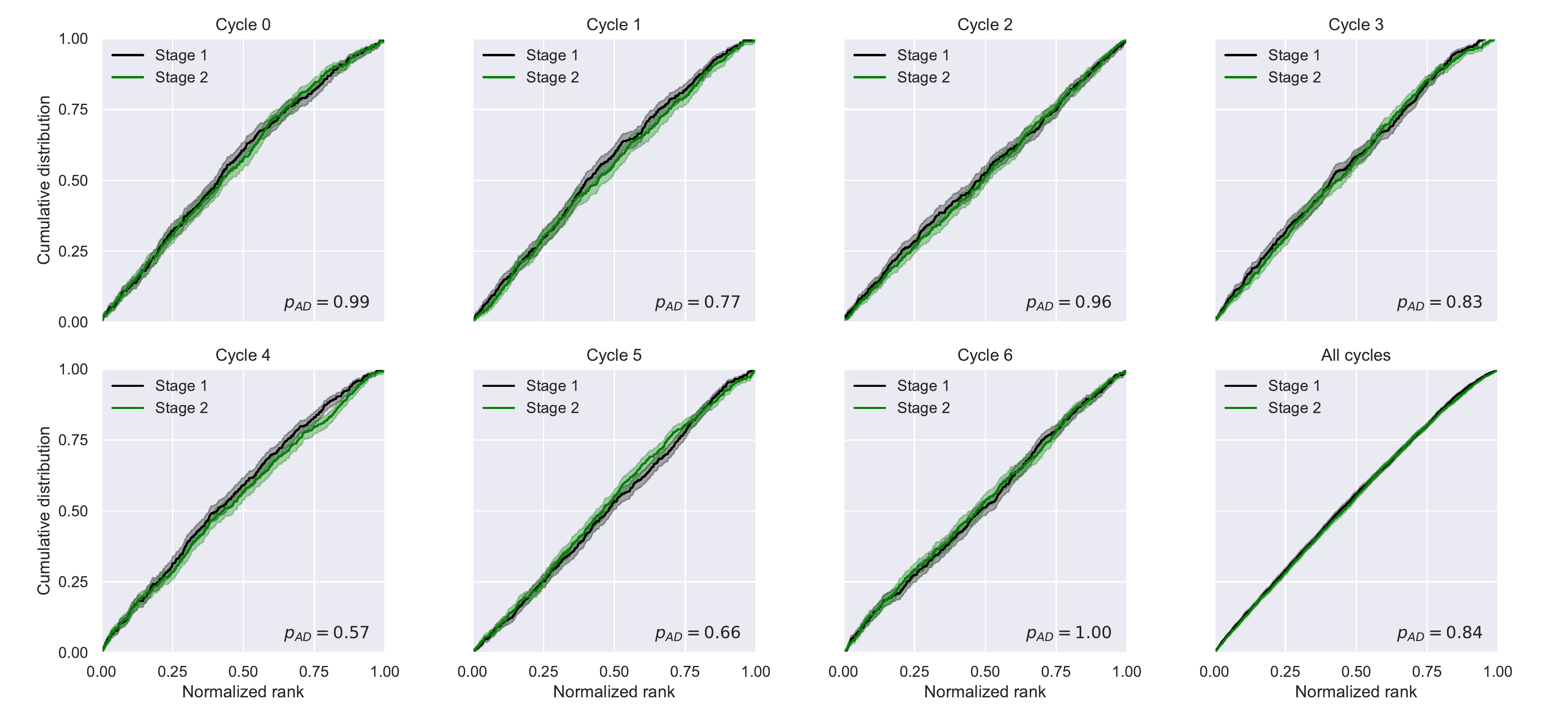}
\caption{
  \label{fig:ad_na_nontriaged}
  Cumulative distribution of Stage 1 and Stage 2 proposal ranks for 
  non-triaged proposals led by an North American PI in Cycles 0--6 and for all cycles combined.
}
\end{figure}

Figures~\ref{fig:ad_cl_nontriaged}, \ref{fig:ad_ea_nontriaged},
\ref{fig:ad_eu_nontriaged}, and \ref{fig:ad_na_nontriaged} compare the Stage 1
and Stage 2 proposal ranks in all seven ALMA cycles for PIs from Chile, East
Asia, Europe, and North America, respectively. Each figure also includes a plot
that combines the results from all cycles. While none of the differences
between the Stage 1 and Stage 2 ranks are significant in any given region or
cycle, some tendencies are seen. Proposals from East Asia tend to be rated
better as a result of the face-to-face discussions. This was most notable in
Cycle 3 and to a less extent in Cycles 1, 4, and 6
(Figure~\ref{fig:ad_ea_nontriaged}). In contrast, European PIs
(Figure~\ref{fig:ad_eu_nontriaged}) tended toward lower ranks after the
face-to-face discussions in Cycle~3 (third and fourth quartiles) and Cycle~5
(second and third quartiles). 

Combining all cycles, the probability that the Stage 1 and Stage 2 cumulative
ranks for non-triaged proposals originate from the same population is 0.13 for
Chile, 0.013 for East Asia, 0.07 for Europe, and 0.84 for North America. Thus
there is a marginally significant tendency for the face-to-face discussion to 
improve the rankings of East Asian proposals while negatively impacting the 
overall European proposal ranks.

\begin{figure}
\epsscale{1.19}
\plotone{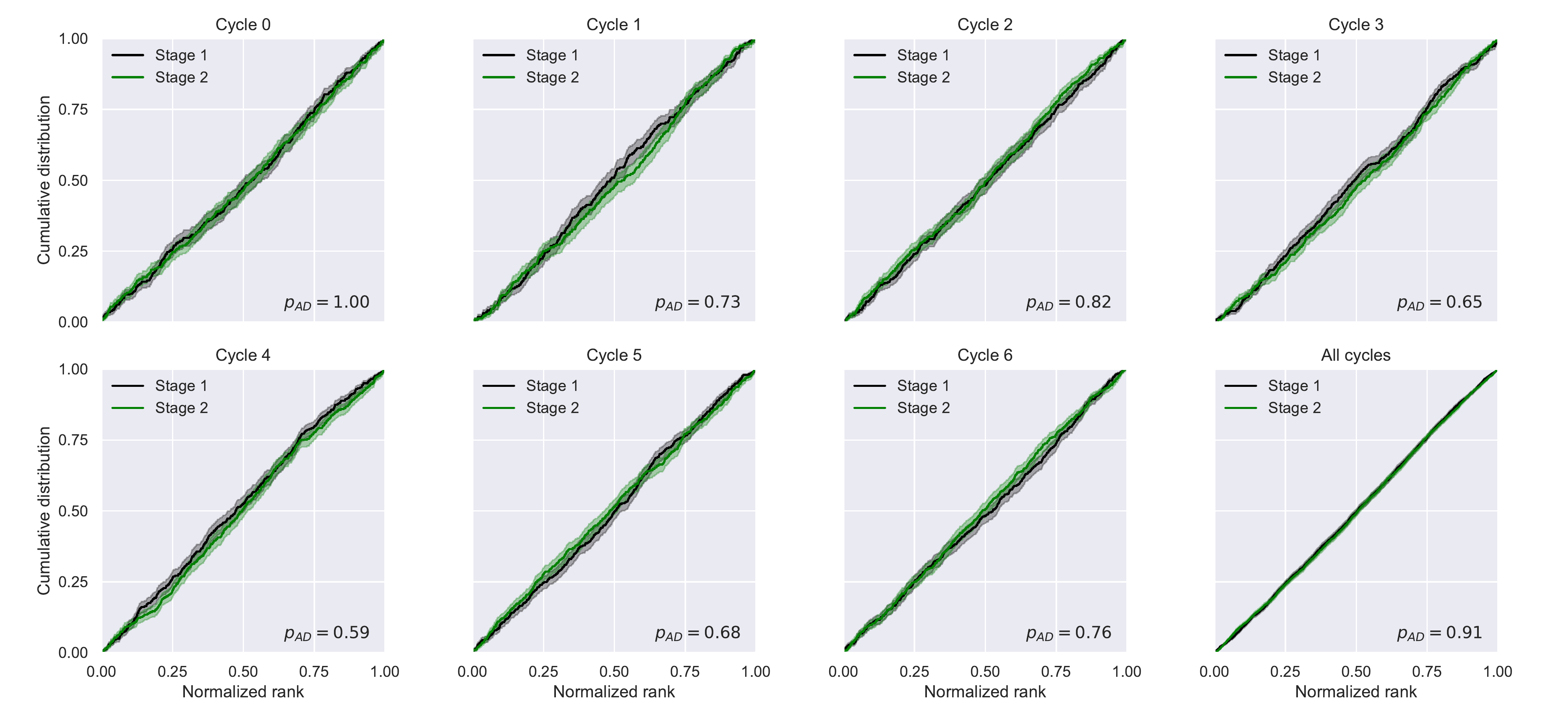}
\caption{
  \label{fig:ad_gender_nontriaged}
  Cumulative distribution of Stage 1 and Stage 2 proposal ranks for 
  non-triaged proposals led by women in Cycles 0--6 and all cycles combined.
}
\end{figure}

Figure~\ref{fig:ad_gender_nontriaged} shows the cumulative Stage 1 and Stage 2
rankings for the non-triaged proposals led by women. No significant or
marginally significant differences between the two distributions are seen in
any cycle. When averaged over all cycles, the probability is 0.91 that the
Stage 1 and Stage 2 ranks for non-triaged proposals led by women PIs share the
same population. Thus the two distributions are indistinguishable. 

In summary, the results shown in
Figures~\ref{fig:ad_experience_nontriaged}-\ref{fig:ad_gender_nontriaged}
indicate that no significant systematics in the proposal rankings are
introduced by the face-to-face review in terms of experience, regional
affiliation, or gender in any given cycle. When averaged over all cycles, East
Asian proposals tend to improve their rankings in the Stage 2 process relative
to Stage 1.

\vspace{0.4in}
\subsection{Proposal priority grades}
\label{grades}

Priority grades for the observing queue are assigned to the proposals by the JAO
based on the Stage 2 rankings and also the share of time per executive. In
recent cycles, balancing the proposal pressure in the various array
configurations, local sidereal time, and weather conditions have also been
considered. Thus the systematic in the proposal rankings by region is partially
compensated for when assigning priority grades by ensuring each region obtains
their pre-determined share of time. However, any systematics with gender in the
proposal rankings are not considered when assigning grades.

Similar to the analysis of the gender distribution of triaged proposals, 
demographics need to be considered when comparing the acceptance rate of
proposals led by men and women. The expected acceptance rate, where an 
``accepted'' proposal is defined as receiving a priority grade of A or B, was
computed as 
\begin{equation} f_{AB,expected} =
\frac{\sum\limits_{Region}\sum\limits_{Experience}\sum\limits_{Category}
f_g(R,E,C) N_{AB}(R,E,C)}
{\sum\limits_{Region}\sum\limits_{Experience}\sum\limits_{Category}
f_g(R,E,C) N_{tot}(R,E,C)}
\end{equation} 
\noindent where $f_g(R,E,C)$ is the fraction of the proposals in the demographic group $(R,E,C)$ that have gender $g$, $N_{AB}(R,E,C)$ is the number of 
proposals awarded grade A or B, and $N_{tot}(R,E,C)$ 
is the total number of submitted proposals.

Table~\ref{tbl:acceptance} lists the proposal acceptance rate by cycle for 
female and male PIs and the expected acceptance rate based on demographics. 
Because of the demographics, the overall acceptance rate of female PIs is
expected to be lower than male PIs. This can be attributed primarily to
regional differences in that Europe and North America have the highest regional
fraction of female PIs and high oversubscription rates while Chile and East
Asia have a lower fraction of female PIs and often lower oversubscription
rates. In addition, an increasing fraction of the PIs in East Asia and Europe
are female, and relatively inexperienced PIs have poorer proposal rankings.

\begin{figure}
\epsscale{1.19}
\plotone{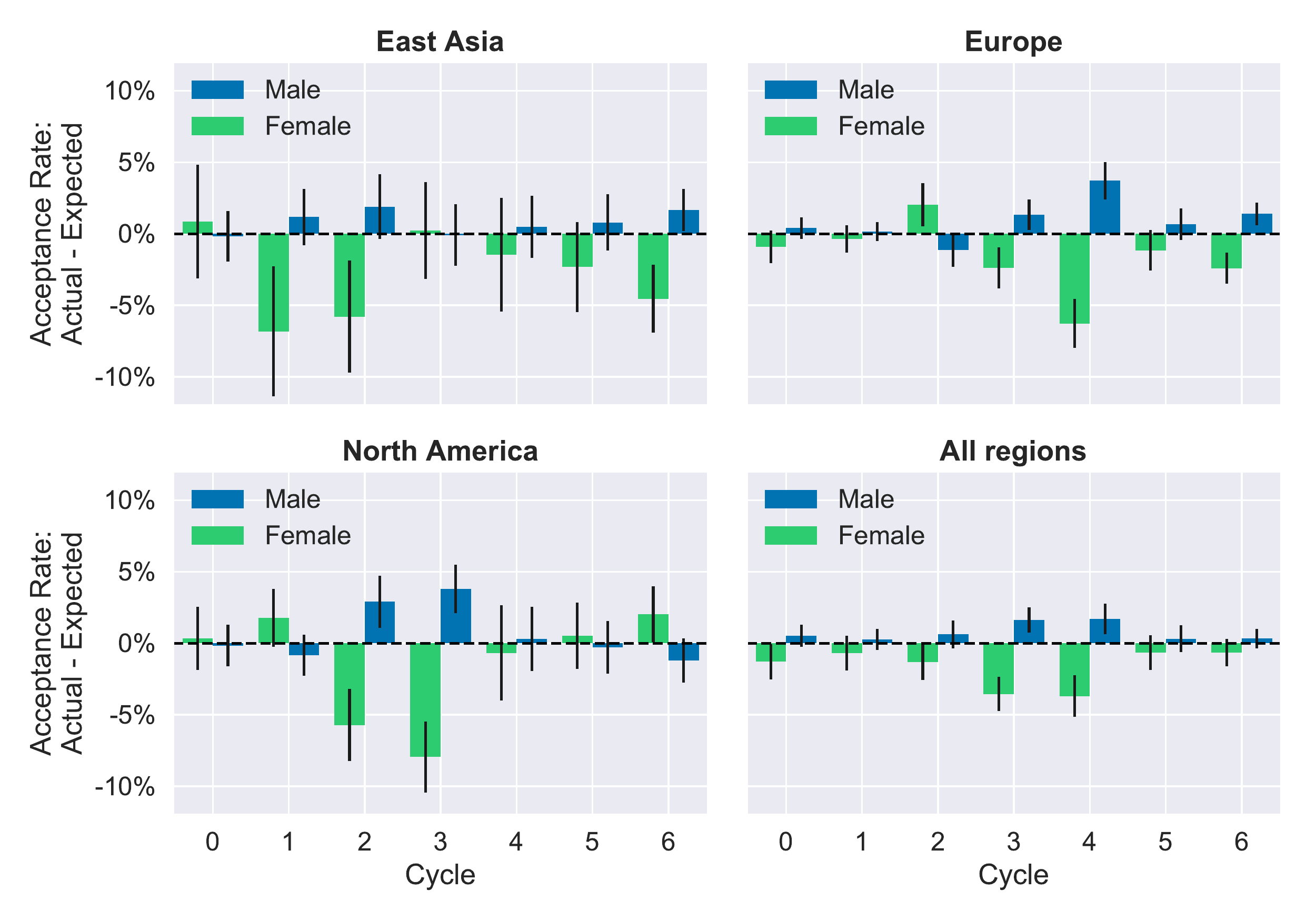}
\caption{
  \label{fig:acceptance}
  The difference between the actual and expected acceptance rate of proposals 
  with female and male PIs by cycle for East Asia, Europe, North America,
  and all regions combined (including Chile and non-ALMA regions). The 
  vertical bars indicate the $1\sigma$ uncertainties.
}
\end{figure}

Figure~\ref{fig:acceptance} plots the difference between the actual and
expected proposal acceptance rate by gender for East Asia, Europe, North
American, and all regions combined, including Chile and non-ALMA regions. Plots
are not shown for Chile and non-ALMA regions since these regions have
relatively few proposals and have large uncertainties. Considering all regions,
female PIs have had a smaller fraction of their proposals assigned priority
Grade A or B than expected in each cycle, although the differences are not 
statistically significant in any given cycle. Conversely, male PIs have had a
higher acceptance rate than expected in each cycle. The difference was largest
is Cycles 3 and 4 and has diminished in Cycles 5 and 6. The Cycle 3 result
was anticipated based on the Stage 1 rankings (see Figure~\ref{fig:ad_gender}).
The Cycle 4 result is more surprising in that the Stage 1 rankings for female
and male PIs in Cycle 4 are indistinguishable. However, in that cycle, female
PIs tended to have lower proposal ranks after the Stage 2 process (see
Figure~\ref{fig:ad_gender_nontriaged}). Even though the difference was not
statistically significant, it was enough to lower the overall acceptance rate.

The trends are similar for individual regions. In East Asia, female PIs have
had a lower acceptance rate than expected in 5 of the 7 cycles. In Europe,
female PIs exceeded the expected acceptance rate in only one cycle. While
female PIs in North America have exceeded expectations in 4 of the 7 cycles,
the cycles where the acceptance rate is below expectations (Cycles 2 and 3 in
particular) are more extreme than when the acceptance rate is higher than
expected.

\begin{deluxetable}{@{\extracolsep{4pt}}ccccc}
\tabletypesize{\scriptsize}
\tablecaption{Acceptance Rate of Proposals\label{tbl:acceptance}}
\tablehead{
   \colhead{Cycle} &
   \multicolumn{2}{c}{Female} &
   \multicolumn{2}{c}{Male}\\
   \cline{2-3}
   \cline{4-5}
   \colhead{} &
   \colhead{$f_{AB}$} &
   \colhead{$f_{AB,expected}$} &
   \colhead{$f_{AB}$} &
   \colhead{$f_{AB,expected}$}
}
\startdata
\multicolumn{5}{c}{\it Chile}\\
   0  &   10.0\% &  44.3\% $\pm$ 16.2\% &   56.0\% &  42.3\% $\pm$  9.9\% \\
   1  &   37.5\% &  46.3\% $\pm$ 13.6\% &   34.7\% &  31.8\% $\pm$  5.5\% \\
   2  &   33.3\% &  23.0\% $\pm$  5.5\% &   36.6\% &  40.1\% $\pm$  5.2\% \\
   3  &   23.5\% &  25.6\% $\pm$  8.1\% &   39.8\% &  39.4\% $\pm$  4.4\% \\
   4  &   15.8\% &  28.6\% $\pm$  7.1\% &   44.2\% &  41.0\% $\pm$  5.5\% \\
   5  &   61.1\% &  55.3\% $\pm$ 13.8\% &   52.9\% &  54.3\% $\pm$  7.1\% \\
   6  &   42.9\% &  29.7\% $\pm$  8.2\% &   31.8\% &  35.0\% $\pm$  4.4\% \\
\multicolumn{5}{c}{\it East Asia}\\
   0  &   22.6\% &  21.7\% $\pm$  4.0\% &   21.1\% &  21.2\% $\pm$  1.8\% \\
   1  &   16.1\% &  23.0\% $\pm$  4.5\% &   24.9\% &  23.7\% $\pm$  2.0\% \\
   2  &   23.9\% &  29.7\% $\pm$  3.9\% &   32.7\% &  30.8\% $\pm$  2.3\% \\
   3  &   28.2\% &  28.0\% $\pm$  3.4\% &   29.7\% &  29.8\% $\pm$  2.2\% \\
   4  &   31.3\% &  32.8\% $\pm$  4.0\% &   32.4\% &  31.9\% $\pm$  2.2\% \\
   5  &   21.4\% &  23.7\% $\pm$  3.2\% &   28.0\% &  27.3\% $\pm$  2.0\% \\
   6  &   14.3\% &  18.8\% $\pm$  2.4\% &   22.3\% &  20.6\% $\pm$  1.5\% \\
\multicolumn{5}{c}{\it Europe}\\
   0  &   11.5\% &  12.4\% $\pm$  1.1\% &   12.9\% &  12.5\% $\pm$  0.8\% \\
   1  &   10.2\% &  10.6\% $\pm$  0.9\% &   11.2\% &  11.1\% $\pm$  0.6\% \\
   2  &   21.9\% &  19.9\% $\pm$  1.5\% &   19.9\% &  21.0\% $\pm$  1.2\% \\
   3  &   18.1\% &  20.4\% $\pm$  1.4\% &   21.7\% &  20.4\% $\pm$  1.1\% \\
   4  &   18.7\% &  24.9\% $\pm$  1.7\% &   28.3\% &  24.6\% $\pm$  1.3\% \\
   5  &   20.0\% &  21.2\% $\pm$  1.4\% &   21.9\% &  21.3\% $\pm$  1.1\% \\
   6  &   14.1\% &  16.5\% $\pm$  1.1\% &   16.9\% &  15.5\% $\pm$  0.8\% \\
\multicolumn{5}{c}{\it North America}\\
   0  &   21.1\% &  20.8\% $\pm$  2.2\% &   20.0\% &  20.2\% $\pm$  1.5\% \\
   1  &   22.0\% &  20.2\% $\pm$  2.0\% &   19.7\% &  20.5\% $\pm$  1.4\% \\
   2  &   21.4\% &  27.1\% $\pm$  2.5\% &   31.2\% &  28.3\% $\pm$  1.8\% \\
   3  &   20.7\% &  28.6\% $\pm$  2.5\% &   32.2\% &  28.4\% $\pm$  1.7\% \\
   4  &   35.9\% &  36.6\% $\pm$  3.3\% &   37.0\% &  36.7\% $\pm$  2.2\% \\
   5  &   28.7\% &  28.2\% $\pm$  2.3\% &   29.6\% &  29.9\% $\pm$  1.8\% \\
   6  &   25.8\% &  23.7\% $\pm$  1.9\% &   24.0\% &  25.2\% $\pm$  1.5\% \\
\multicolumn{5}{c}{\it Other}\\
   0  &   20.0\% &   7.3\% $\pm$  3.6\% &    6.2\% &  10.2\% $\pm$  3.5\% \\
   1  &   20.0\% &  20.0\% $\pm$ 14.1\% &    4.0\% &   4.0\% $\pm$  2.8\% \\
   2  &    0.0\% &   8.3\% $\pm$  5.9\% &   13.8\% &  12.1\% $\pm$  6.1\% \\
   3  &   20.0\% &  19.6\% $\pm$  6.9\% &   13.3\% &  13.6\% $\pm$  4.8\% \\
   4  &   40.0\% &  29.0\% $\pm$  9.5\% &   26.7\% &  32.2\% $\pm$  8.1\% \\
   5  &    9.1\% &  14.4\% $\pm$  7.5\% &   13.5\% &  11.9\% $\pm$  3.0\% \\
   6  &   30.8\% &  25.6\% $\pm$ 10.7\% &    2.3\% &   3.9\% $\pm$  1.6\% \\
\multicolumn{5}{c}{\it All regions}\\
   0  &   16.3\% &  17.6\% $\pm$  1.2\% &   18.3\% &  17.8\% $\pm$  0.8\% \\
   1  &   16.6\% &  17.2\% $\pm$  1.2\% &   17.7\% &  17.5\% $\pm$  0.7\% \\
   2  &   22.4\% &  23.7\% $\pm$  1.3\% &   27.0\% &  26.4\% $\pm$  1.0\% \\
   3  &   20.7\% &  24.2\% $\pm$  1.2\% &   27.8\% &  26.1\% $\pm$  0.9\% \\
   4  &   25.9\% &  29.6\% $\pm$  1.5\% &   32.8\% &  31.1\% $\pm$  1.1\% \\
   5  &   24.2\% &  24.8\% $\pm$  1.2\% &   27.2\% &  26.8\% $\pm$  0.9\% \\
   6  &   19.2\% &  19.8\% $\pm$  1.0\% &   20.5\% &  20.2\% $\pm$  0.7\% \\
\enddata
\tablecomments{The table lists the fraction of proposals assigned priority 
grade A or B ($f_{AB}$) with a female or male PI. Also listed is the expected 
fraction ($f_{AB,expected}$) given the demographics of the accepted proposals, 
as described in the text.
}
\end{deluxetable}

\vspace{0.2in}
\section{Summary and Conclusions}
\label{sec:summary}

An analysis is presented of the outcomes of the proposal review process in ALMA
Cycles 0-6 to identify systematics in the proposal rankings that may signify
potential bias with respect to the experience level, regional affiliation, or
gender of the PI. The analysis was conducted for both the Stage 1 rankings,
which are based on the preliminary scores of the reviewers, and the Stage 2
rankings, which are based on the face-to-face panel discussions. The results
show that systematics are introduced primarily in the Stage 1 process and not
the face-to-face discussion.

One significant trend is that PIs who submit proposals every cycle have better
proposal ranks than PIs who have submitted proposals for the first time. The
trend is also present within intermediate levels of experience. One should not
expect a completely random correlation of the proposal ranks with experience
level since many expert PIs will have a detailed understanding on how best to
use the ALMA and it would not be surprising if they can write compelling
proposals. Also, PIs who resubmit previously declined proposals have feedback
from the reviewers on how to improve the proposal. The question remains,
however, to what degree reviewers give experienced PIs leeway in the proposal
review on perceived prestige, but the data in hand cannot address that question
directly.

A second significant trend is that proposals submitted by PIs from North
America and Europe have better ranked proposals than PIs from East Asia
and Chile, although a small improvement in the proposal rankings for East
Asia are observed as a result of the Stage 2 discussions. This trend is not
unique to ALMA. \citet{Reid14} reports that PIs from Europe and North America
have a higher success rate on proposals submitted to the Hubble Space Telescope
(HST) than PIs from the rest of the world. 

While the origin of the trend in the ALMA proposal rankings with region is
unclear, one speculation is that it can be attributed to differences in the
proficiency in using the English language. ALMA proposals are mandated to be
written in English, which is a second language for a large fraction of ALMA
PIs. However, the proficiency level in English likely varies between regions
and may cause reviewers to penalize proposals that are not as well written even
if the underlying science is strong. Stylistic differences in how a proposal is
structured may also potentially exist between regions. \citet{Hall76}
introduced the concept of high- and low-context communication and indicated
communication styles can vary between countries. High-context cultures rely
heavily on non-verbal methods to convey information, while low-context cultures
communicate information primarily through language. \citet{Hall76} indicated
that Japan has high-context communication while the United States is low
context. Since approximately two thirds of the ALMA reviewers are from Europe
and North America, the proposal rankings could reflect preference toward
proposal styles from those regions or that low-context styles are more suitable
to proposal peer review. It is tempting to conclude that the improvement in the
scientific rankings of East Asian proposals during the Stage 2 face-to-face
discussions is a result of overcoming potential language biases or stylistic
differences to select the best science, but that remains speculation for now.

As noted by \citet{Lonsdale16}, male PIs tend to have better ALMA proposal
ranks than female PIs in Cycles 2-4. This was most apparent in Cycle 3, but
since then, no measurable differences exist in the cumulative Stage 1
proposal rankings between women and men in Cycles 4-6 even when individual
cycles are combined. Nonetheless, the proposal acceptance rates, which
ultimately reflect what is scheduled on the telescope, show a similar trend
as HST \citep{Reid14} and ESO \citep{Patat16} in that proposals with female PIs
have had a lower success rate in receiving telescope time than proposals
with male PIs, even after accounting for demographic differences by region,
experience, and science category. The difference between the actual and
expected proposal acceptance rate for women is not significant in any given
cycle, but is present in each cycle when considering all regions combined.
Whether the systematic differences in the acceptance rate represents a bias 
in the review process or an unaccounted for demographic difference between 
women and men \citep[e.g., seniority; c.f.][]{Patat16} is unclear.

Identifying the underlying causes of the systematics in the proposal rankings
is difficult. Multiple factors could plausibly contribute to the observed
trends and limited ancillary demographic data (e.g., seniority of the PI) is
present to test various hypotheses. A deeper and more sophisticated analysis
than presented here may clarify some of the causes, including understanding to
what extent any systematics depend on the regional affiliation of the proposal
reviewers. Establishing an objective measure of any stylistic differences in
the proposals by region in particular could yield important insights into the
region-based systematics.

In Cycle 7, ALMA took steps to reduce the impact of potential biases in the
review process. The investigators were listed in random order on the proposal
coversheet such that reviewers will know the members of the proposal team but
not the identity of the PI. Also, first names were listed only by the first
initial so that the gender cannot be readily inferred. These steps are expected
to reduce biases that may be triggered by simply knowing the name of the PI,
but other systematics identified here could remain. For example, if the
differences in proposal ranks by region are caused by language or style, the
systematics in the proposal rankings by region should not change.

These steps to modify the proposal coversheet follow those taken by the HST
after \citet{Reid14} identified a small but persistent effect where the
acceptance rate of HST proposals was lower for women (19\% on average) than men
(23\% on average). Similar steps have recently been taken at ESO.
Interestingly, the difference in the acceptance rate of HST proposals by gender
persisted even after anonymizing the PIs. Only after the list of investigators
was completed hidden from the reviewers in a double-anonymous review did the
acceptance rate for proposals led by women exceed that of men
\citep{Strolger19}. ALMA is following the HST experience and is considering
implementing a double-anonymous review in future cycles.

\acknowledgments
{
I am grateful to Andrea Corvillon for providing the proposal data used in
this analysis. C. Lonsdale kindly provided her tabulation of genders for ALMA
PIs and assisted in updating the information. D. Iono also helped identify many
of the gender demographics for East Asian PIs. I also thank S. Dougherty, G.
Mathys, M. Fukagawa, J. Greaves, L. Barcos, F. Schwab, L. Ball, D. Balser, and
the anonymous referee for comments on the manuscript.
}

\software{Astropy \citep{astropy:2018}, SciPy \citep{Jones01}, kSamples, \citep{Scholz19}}

\appendix

\section{Proposal Acceptance Rate for ALMA Reviewers}

\citet{Greaves18} analyzed the proposal statistics for an anonymous observatory
using published lists of reviewers and accepted proposals. Private
communication with J. Greaves confirmed that the observatory is ALMA. The
main result of the paper is that the Cycles 2-4 reviewers had a three-fold
increase in the number of proposals accepted while serving on the panel
compared to when they were not serving on a review panel. The inference was
that there is a bias in the review process where reviewers preferentially favor
proposals from the other review participants. Since the proposals submitted by
a given reviewer are generally reviewed in a different panel, the implication
is that reviewers are predisposed toward proposals from reviewers in other
panels. A limitation of this analysis, however, is that \citet{Greaves18} did
not have access to the total number of proposals submitted by the reviewers and
could not compute the fraction of submitted proposals that were accepted.

This appendix investigates the result from \citet{Greaves18} by analyzing both
the accepted and rejected proposals to examine the proposal acceptance rate of
the reviewers and not just the number of accepted proposals.
Table~\ref{tbl:reviewers} presents the number of proposals submitted and
accepted by cycle for reviewers while they served on the ALMA review panels 
in any cycle and when they were not on the review panels. Most reviewers serve
for three consecutive cycles. For example, the Cycle 0 reviewers also typically
served on the review panels in Cycles 1 and 2. The Cycle 0 reviewers submitted
130 proposals while serving on a panel in any cycle, and 48 were awarded Grade
A or B for an overall acceptance fraction of 36.9\%. By comparison, when the
Cycle 0 reviewers were not serving on a panel, they have submitted 187
proposals and 60 have been awarded Grade A or B for an acceptance fraction of
32.1\%. The uncertainties in each of the acceptance fractions is $\sim$4\%, and
therefore the difference in the acceptance rate between when on and off a panel
is not significant.

Examining all cycles, reviewers serving on the panels in Cycles 0-4 tend to
have a higher acceptance rate than when they are not on the panels. The trend
reversed in Cycles 5 and 6 when the reviewers had a lower acceptance rate. When
measured over all cycles, the acceptance of reviewers when serving on the panel
is 36.9\%$\pm$1.1 compared to 34.5\%$\pm$0.9 when not present on the panels.
None of these differences are statistically significant. Thus no discernible
bias is present that favors ALMA reviewers when they are present on the panels
versus when they are off the panels.

The second column in Table~\ref{tbl:reviewers} shows the fraction of proposals
accepted for all PIs in that cycle. The acceptance rate for all PIs varies
between 17.4\% and 30.2\% depending on the oversubscription rate in a given
cycle. Table~\ref{tbl:reviewers} shows that reviewers consistently have a
higher acceptance rate than the overall average by as much as a factor of two
in some cycles. However, this trend is present when the reviewers are not 
on the panels. In conclusion, the astronomers selected to serve on the ALMA
review panels do have an higher acceptance rate than the typical PI, but this 
is true even when they are not serving on a panel.

\begin{deluxetable}{@{\extracolsep{4pt}}cccccccc}
\tabletypesize{\normalsize}
\tablecaption{Proposal Statistics for ALMA Reviewers\label{tbl:reviewers}}
\tablehead{
   \colhead{Cycle} &
   \multicolumn{1}{c}{All PIs} &
   \multicolumn{3}{c}{Reviewers serving on a panel} &
   \multicolumn{3}{c}{Reviewers not serving on a panel}\\
 \cline{2-2}
 \cline{3-5}
 \cline{6-8}
 \colhead{} &
 \colhead{Acceptance} &
 \colhead{Submitted} &
 \colhead{Accepted} &
 \colhead{Acceptance} &
 \colhead{Submitted} &
 \colhead{Accepted} &
 \colhead{Acceptance}\\
 \colhead{} &
 \colhead{fraction} &
 \colhead{} &
 \colhead{} &
 \colhead{fraction} &
 \colhead{} &
 \colhead{} &
 \colhead{fraction}
}
\startdata
  0 & $17.7\%^{+1.3}_{-1.2}$ &  130 &   48 & $36.9\%^{+4.3}_{-4.1}$ &  187 &   60 & $32.1\%^{+3.5}_{-3.3}$\\
  1 & $17.4\%^{+1.2}_{-1.1}$ &  211 &   79 & $37.4\%^{+3.4}_{-3.3}$ &  279 &   93 & $33.3\%^{+2.9}_{-2.8}$\\
  2 & $25.6\%^{+1.2}_{-1.2}$ &  233 &   92 & $39.5\%^{+3.2}_{-3.2}$ &  292 &   97 & $33.2\%^{+2.8}_{-2.7}$\\
  3 & $25.5\%^{+1.1}_{-1.1}$ &  234 &   95 & $40.6\%^{+3.2}_{-3.2}$ &  271 &   95 & $35.1\%^{+2.9}_{-2.8}$\\
  4 & $30.2\%^{+1.2}_{-1.1}$ &  350 &  128 & $36.6\%^{+2.6}_{-2.5}$ &  497 &  165 & $33.2\%^{+2.1}_{-2.1}$\\
  5 & $26.1\%^{+1.1}_{-1.1}$ &  343 &  118 & $34.4\%^{+2.6}_{-2.5}$ &  565 &  197 & $34.9\%^{+2.0}_{-2.0}$\\
  6 & $20.1\%^{+1.0}_{-0.9}$ &  348 &  122 & $35.1\%^{+2.6}_{-2.5}$ &  704 &  258 & $36.6\%^{+1.8}_{-1.8}$\\
All & $23.7\%^{+0.4}_{-0.4}$ & 1849 &  682 & $36.9\%^{+1.1}_{-1.1}$ & 2795 &  965 & $34.5\%^{+0.9}_{-0.9}$\\
\enddata
\end{deluxetable}

\end{document}